\documentclass[aps,prb,twocolumn,superscriptaddress,showpacs, nofootinbib]{revtex4-2}
\usepackage{graphicx}% Include figure files
\usepackage{float}
\usepackage{dcolumn}% Align table columns on decimal point
\usepackage{float}
\usepackage{bm}% bold math
\usepackage{mathrsfs}
\usepackage{fullpage}
\usepackage{ulem}
\usepackage[per-mode = fraction]{siunitx}
\usepackage{hyperref}
\usepackage{url}
\hypersetup{
    colorlinks=true,
    linkcolor=blue,
    filecolor=magenta,      
    urlcolor=cyan,
    citecolor=magenta,
}
\usepackage{xcolor}
\usepackage{soul}
\usepackage{amsthm,amssymb}
\usepackage{mathtools}
\usepackage{comment}
\usepackage{physics}

\begin{document}

%\title{Asymmetric transport and nonreciprocal superconductivity in the fractional-order Ginzburg–Landau theory}

%\title{The superconducting diode effect in the fractional-order Ginzburg–Landau theory}
%\title{Superconducting diode effect in fractal superconductors: fractional-order Ginzburg–Landau Josephson junctions}
\title{Superconducting diode effect in fractal superconductors: fractional-order Ginzburg–Landau theory for Josephson junctions}

\author{Yuriy Yerin}
\affiliation{Istituto di Struttura della Materia of the National Research Council, via Salaria Km 29.3, I-00016 Monterotondo Stazione, Italy}
\affiliation{Department of Computer Engineering and Center of Excellence of Superconductivity Research, Ankara University, Ankara, 06100, Turkey}
\author{Iman Askerzade}
\affiliation{Department of Computer Engineering and Center of Excellence of Superconductivity Research, Ankara University, Ankara, 06100, Turkey}
\affiliation{Institute of Physics, H.Cavid 33, Baku, AZ1143, Azerbaijan}
\affiliation{Center for Theoretical Physics, Khazar University, 41 Mehseti Street, Baku, AZ1096, Azerbaijan}

\date{\today}

\begin{abstract}

We develop a fractional-order Ginzburg–Landau (GL) framework for nonreciprocal superconducting transport in Josephson junctions formed by fractal superconductors or superconducting media with nonlocal correlations, separated by a noncentrosymmetric normal layer. We show that nonreciprocity and the superconducting diode effect arise from the interplay between the Lifshitz invariant and fractional kinetics, with the latter serving as an effective, symmetry-consistent representation of fractal geometry and finite-range memory. Two complementary approaches are pursued.  In a fractional integral (measure-weighted) GL formulation, spatial integration on a fractal space yields analytic solutions and reveals how rectification scales with the dimensionality of the fractal media and the strength of the Lifshitz-like drift.  In a fractional derivative-based formulation derived via the Agrawal variational principle with left/right Caputo operators, we obtain a gauge-invariant free energy, the corresponding GL equations, and a current density. We use fractional orders as effective parameters that represent nonlocal and memory effects induced by fractal microstructure; they are not in one-to-one correspondence with a specific fractal geometry. Within a two-mode plane-wave approximation we derive a compact current–phase relation and an expression for the diode efficiency, and we map the rectification amplitude across the fractional kinetic and the Lifshitz/memory order. An exact single-sided solution in terms of Prabhakar functions (generalized Mittag–Leffler) further confirms robust, tunable nonreciprocity, including a near-ideal diode response. This identifies a pathway to near-perfect superconducting diodes by engineering fractal (fractional-kinetic) transport achieved by tuning the fractional orders and Lifshitz strength without invoking magnetic fields or geometric ratchets. In the integer limit of local kinetics and Lifshitz-like drift, both constructions reduce to the standard $\varphi_0$-shifted Josephson relation, embedding the conventional result within a unified fractional GL phenomenology.
\end{abstract}

\maketitle

\section{Introduction}

Nonreciprocity is a recurring theme across many branches of physics. Systems that lack inversion or time-reversal symmetry can exhibit physical properties and characteristics that differ under reversal of current, momentum, or wavevector. Examples range from optical isolators \cite{LeiBi2011, Fang11, XiaoLin2014} and nonreciprocal metamaterials \cite{Brandenbourger2019, Buddhiraju2020, Shaat2020} to magnetoelectric materials \cite{Keller2012, Khandekar_2020, Keller2024} and spin–orbit-coupled conductors \cite{Tokura2018, YanLi2021, Pan2022}. 

In superconductivity, nonreciprocity manifests most prominently through the transport properties in the form of the diode effect, i.e. the critical current depends on the direction of supercurrent flow. This phenomenon has recently attracted intense interest because it combines  dissipationless transport with rectification functionality in both conventional and unconventional superconductors and a wide variety of hybrid superconducting systems \cite{ModernAspects2017, Nadeem2023, review2025, InglaAynes2025}. For now these materials and hybrid structures include a broad family of non-centrosymmetric \cite{WakatsukiSciAdv2017, and20, Zhang2020}, chiral  \cite{Zinkl} patterned  \cite{Lyu2021, Osin2024}, high-temperature \cite{Ghosh2024}, multiband \cite{Yerin_diode, Yerin_SQUID} superconductors, two-dimensional electron gases, semiconductors \cite{Itahashi20, mondal2025}, Dirac materials \cite{Dirac_diode}, superconductor-magnet hybrids \cite{wu22, Narita2022, Sun_altermagnet}, Josephson junctions that incorporate magnetic atoms \cite{tra23}, twisted graphene systems \cite{lin22, Scammell_2022}, and topological insulators \cite{BoLu2022, BoLu2023}. Numerous physical scenarios and mechanisms have been exploited to achieve  nonreciprocal transport properties in these systems, based on finite Cooper pair momentum \cite{pal22,yuan22}, helical phases \cite{Edelstein95,Ilic22,dai22,he22,Turini22}, anisotropic and multi-component structure of the order parameter \cite{askerzade_2012, askerzade_2015, Yerin2014, Kiyko, Yerin_review}, magnetic texture and magnetization gradients \cite{Cayao_Majorana, roig2024}, the presence of screening currents \cite{hou23,sun23}, and supercurrents connected with self-induced fields \cite{kras97,GolodNatComms2022}.

As a result, a variety of theoretical approaches have been applied to describe the superconducting diode effect, reflecting its microscopic richness and the interplay of symmetry breaking, spin-orbit coupling, and electronic correlations. On the microscopic level, the formalism  based on the Bogoliubov-de Gennes equations, Green's functions, or diagrammatic perturbation theory provides a detailed account of quasiparticle spectra and current-phase relations, but at the cost of considerable complexity. In contrast, the Ginzburg-Landau (GL) phenomenological theory offers a simpler and more transparent framework \cite{He2022}, where the essential physics of nonreciprocal transport is encoded in symmetry-allowed gradient terms of the order parameter such as Lifshitz invariants. Although the GL approach necessarily sacrifices microscopic detail, its analytical tractability makes it particularly suitable for exploring general mechanisms, establishing connections with symmetry groups, and providing qualitative guidance for experimental observations.

In recent years, there has been interest in extending the GL theory itself by incorporating concepts of fractional calculus operating by fractional derivatives and
integrals \cite{Milovanov2005, Ali2023}.  The fractional-order GL theory provides a natural extension of the conventional GL framework when the superconducting condensate exhibits anomalous spatial correlations, fractal geometries, or nonlocal interactions. Unlike the standard GL equation, where the kinetic term involves a second-order Laplacian (or double Laplacian operator for the Fulde-Ferrell-Larkin-Ovchinnikov state), the fractional formulation replaces this operator by a fractional derivative or so-called weighted derivative that interpolates between local and nonlocal correlations. This modification directly encodes the presence of memory effects, or fractal effective dimensionalities, all of which are increasingly relevant for certain superconducting materials. For example, low-dimensional superconductors, inhomogeneous thin films, and systems with engineered nanostructures often display power-law correlations and non-exponential healing of the order parameter, behaviors that may be naturally captured within a fractional GL approach.
To be more specific, fractional calculus within the framework of the GL theory generalized to the case of fractional or fractal dimensions has already been applied to describe the non-local effects and an effective coherence length  in a bulk superconductor, the structure of the Abrikosov vortex, the non-periodic vortex matter in an anisotropic superconducting medium, and the emergence of type-III superconductivity \cite{Anukool1, Anukool2, Aguilar2019}.

There are several reasons why fractional calculus may also be useful in describing the nonreciprocal transport properties in superconductors, and in particular the Josephson effect. First, as mentioned before, fractional operators encode nonlocality in space, reflecting the fact that superconducting correlations extend over finite coherence lengths and can involve long-range coupling in unconventional or disordered Josephson systems \cite{Kruchinin}. Second, they naturally capture power-law relaxation and anomalous scaling, which arise in vortex dynamics, proximity effects, and systems with fractal microstructures. Third, the Josephson effect itself is inherently nonlocal: the current–phase relation reflects the phase difference throughout the weak link rather than a purely local gradient. Fractional GL operators provide a compact and physically motivated way to model such distributed effects. In addition, they also allow for the capture of deviations from the simple sinusoidal current–phase relation, such as higher harmonics and anomalous current-phase relations observed in unconventional or strongly disordered Josephson junctions. From this perspective, the fractional-order framework is not merely a mathematical curiosity and Gedankenexperiment, but a physically justified extension that may unify a variety of observed nonlocal and anomalous Josephson effects. Therefore, within the context of the superconducting diode effect, fractional derivatives offer a powerful tool for describing inversion-symmetry breaking and nonreciprocal transport beyond the standard GL model with integer dimensionality.

Motivated by these developments, in this paper we explore the superconducting diode effect within a fractional-order GL theory in a Josephson junction between fractal superconductors. By introducing fractional calculus formalism into the GL theory, we study the interplay between nonlocal order-parameter correlations and the Lifshitz term manifested noncentrosymmetrical superconductivity. We analyze the resulting current–phase relations, extract the efficiency of the diode, and track the dependence of the rectification amplitude on the fractional order (fractal dimensionality) of the derivative.  Throughout, we treat fractional orders  as effective transport parameters capturing anomalous, scale-free behavior generated by fractal microstructure; they serve as surrogate descriptors rather than unique identifiers of any given fractal. Our results show that fractionalization can qualitatively modify the nonreciprocal response of a Josepshon junction, offering a new route to tune and significantly enhance the diode effect in a wide class of superconducting systems.

\section{Model and basic formalism}

\begin{figure}[h]
\includegraphics[width=1\linewidth]{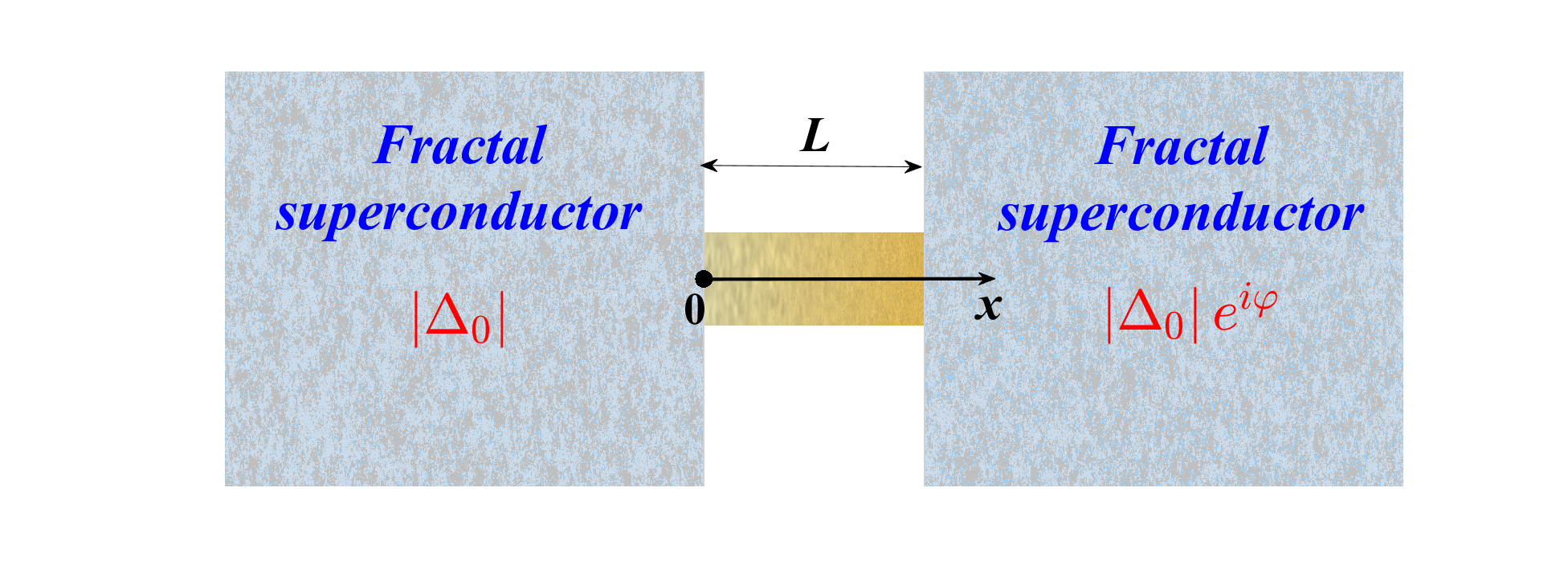}
\caption{Schematic illustration of the Josephson junction as a weak link between two fractal superconductors connected by a filament with length $L$. The fractality of superconducting banks is depicted in the form of disordered granular structures. The normal non-centrosymmetric material of the filament is emphasized by coarse-to-fine shading.}
\label{model}
\end{figure}

The Josephson junction is modeled as a weak link of length $L$ formed by a normal normal with the lack of inversion symmetry between two bulk superconductors with a fractal structure (Fig. \ref{model}). Also, we consider that the temperature is above the critical temperature of the material of the weak link, and the superconducting order parameter is induced only by the superconducting banks.

We employ two complementary GL approaches based on the fractional calculus formalism. The first is formulated in terms of fractional integrals for the GL functional, which provide an effective description of media with fractal microstructure. The second is based on the Agrawal variational principle \cite{Agrawal} and Caputo fractional derivatives \cite{Caputo}, which allows us to directly generalize the GL equations and the expression for the current density by introducing left/right-sided differential operators. These two theoretical tools are mathematically related to each other and together they provide a framework for description of superconducting transport properties and prediction of the diode effect in a Josephson junction between fractal superconductors separated by a noncentrosymmetric metal.

\section{Fractional generalization of the GL functional}

We introduce fractional generalization of the GL functional with fixed ${\bf{A}} = 0$, i.e. there is no magnetic field:
\begin{widetext}
\begin{equation}
F [\Delta] = {F_0} + \int\limits_\Omega  {d{V_d}\left[ {a {{\left| \Delta(x)  \right|}^2} + \frac{1}{2} b {{\left| \Delta(x)  \right|}^4} + \gamma {{\left| \Delta'(x) \right|}^2} - \frac{1}{2}i\varepsilon w(x) \left( {\Delta(x) \Delta'^*(x) - {\Delta(x) ^ * }\Delta'(x)} \right)} \right]},
\label{GL_free_energy}
\end{equation}
\end{widetext}
where $\Delta$ is the superconducting order parameter, $a<0$, $b>0$, $\gamma>0$ are phenomenological coefficients, $F_0$ is a free energy of the normal state. In what follows, we work with a reduced 1D GL functional and replace the full vector structure of the Rashba spin–orbit coupling and the exchange field $h$  by a single effective phenomenological coefficient $\varepsilon$, so that the corresponding Lifshitz invariant appears as a scalar antisymmetric gradient term in Eq. \eqref{GL_free_energy} \cite{Edelstein_1996, Kaur, AGTERBERG200313, Buzdin, agterberg2011}. 

Integration is performed in the Josephson junction space $\Omega$ with the element of the effective fractal volume $dV_d$ defined in terms of the Riemann-Liouville fractional integral:
\begin{equation}
d{V_d} = \frac{{{{ x }^{d - 1}}}}{{\Gamma \left( d \right)}} d x,
\label{metric}
\end{equation}
where we set the dimensionality in the interval $0<d \leq 1$ with $\Gamma(d)$ as the Gamma function. In the following, to avoid confusion with the dimension of the space $d$, we denote derivatives with respect to the coordinate $x$ using prime and double prime symbols. 

The purpose of introducing the function $w(x)$ into the GL functional is motivated by the fact that there exist two formally consistent ways of incorporating the Lifshitz invariant term into Eq. (\ref{GL_free_energy}):
\begin{equation}
w(x)=
\begin{cases}
1, & \text{"weighted" Lifshitz term},\\
\frac{{\Gamma \left( d \right)}}{{{{\left| x \right|}^{d - 1}}}}, & \text{"unweighted" Lifshitz term}.
\end{cases}
\label{w_func}
\end{equation}

In the so-called "weighted" formulation, following the Ref. \onlinecite{Tarasov}, every term in a GL functional is multiplied by the special anomalous "weight", represented by the factor before the differential $dx$ in Eq. (\ref{metric}). It corresponds to the effective fractal geometry of the superconducting medium. This treatment is mathematically rigorous.

By contrast, the microscopic origin of the Lifshitz term is distinct: it arises from spin-orbit coupling combined with broken inversion symmetry, rather than from the fractal measure itself. In this sense, the coordinate $x$ that appears in the Lifshitz term should be regarded as an effective continuum coordinate that describes the texture of the superconducting phase, not as the fractal
integration variable. For this reason, it is physically justified to treat the Lifshitz invariant as an independent "unweighted" contribution to the GL energy, acting as a non-Hermitian drift term \cite{non-Herm}. In other words, while the "weighted" Lifshitz form is natural from the standpoint of fractional calculus, the "unweighted" Lifshitz term provides a more faithful phenomenological description of the Josephson junction between fractal superconductors where fractality and spin-orbit-induced contribution operate as independent mechanisms. This separation of mechanisms ensures consistency with microscopic derivations of Lifshitz invariants in non-centrosymmetric superconductors, where no spatial "weight" factor accompanies the linear gradient term. 

In any case, for clarity, we treat both common concepts at once by introducing a function $w(x)$ into Eq. (\ref{GL_free_energy}).

After varying the functional Eq. (\ref{GL_free_energy}) together with Eqs. (\ref{metric}) and (\ref{w_func}), one can derive the "weighted" linearized GL equation
\begin{equation}
\begin{aligned}
-& \gamma \Bigl( \Delta''(x) + \frac{d-1}{x} \Delta'(x) \Bigr)
+ i\varepsilon \, \Delta'(x) \\[4pt]
&+ \frac{i\varepsilon (d-1)}{2x} \, \Delta(x)
+ a \, \Delta(x) = 0,
\end{aligned}
\label{GL_eq_lin1}
\end{equation}
and its "unweighted" counterpart
\begin{equation}
 - \gamma \left( {\Delta''(x) + \frac{{d - 1}}{x}  \Delta'(x)} \right) + i\varepsilon  \Delta'(x)  + a  \Delta(x)  = 0.
 \label{GL_eq_lin2}
\end{equation}
Both Eqs. (\ref{GL_eq_lin1}) and (\ref{GL_eq_lin2}) can be reduced to canonical forms of differential equations with the general solutions for a "weighted" representation:
\begin{equation}
\label{OP_wL}
\begin{aligned}
\Delta^{\cal W}(x) &= \Delta_0 \, x^{1-\frac{d}{2}} \,
\exp\!\left( \frac{i \varepsilon x}{2\gamma} \right) \\[6pt]
&\quad\times \Bigl[ {\cal W}_1 \,
J_{\frac{d}{2}-1}(\lambda x) + {\cal W}_2 \,
Y_{\frac{d}{2}-1}(\lambda x) \Bigr],
\end{aligned}
\end{equation}
where $J_{\mu}(x)$ and $Y_{\mu}(x)$ are Bessel functions of the first and second kinds, and for a "unweighted" formulation:
\begin{equation}
\label{OP_uL}
\begin{aligned}
\Delta^{\mathcal{U}}(x) &= \Delta_0 \, x^{-\frac{d-1}{2}} \,
\exp\!\left( \frac{i \varepsilon x}{2\gamma} \right) \\[6pt]
&\quad\times \Bigl[ \mathcal{U}_1 \,
M_{\kappa, \frac{d}{2}-1}(2i\lambda x) 
+ \mathcal{U}_2 \, W_{\kappa, \frac{d}{2}-1}(2i\lambda x) \Bigr],
\end{aligned}
\end{equation}
where $M_{\kappa,\mu}(x)$ and $W_{\kappa,\mu}(x)$ are Whittaker functions, parameter $\kappa  = \frac{{\varepsilon (d - 1) }}{{4 \gamma \lambda }}$ and coefficient $\lambda  = \sqrt { - \frac{a }{\gamma } + \frac{{{\varepsilon ^2}}}{{4{\gamma ^2}}}}$.

Here $\mathcal{W}_1$, $\mathcal{W}_2$ and $\mathcal{U}_1$, $\mathcal{U}_2$ are arbitrary constants in Eqs. (\ref{OP_wL}) and (\ref{OP_uL}), respectively. They can be determined from the Dirichlet boundary conditions for the Josephson junction shown in Figure \ref{model}:
\begin{equation}
\begin{array}{l}
\Delta \left( 0 \right) = \left| {{\Delta _0}} \right|, \Delta \left( L \right) = \left| {{\Delta _0}} \right| e^{i\varphi}.
\end{array}
\label{eq:BC}
\end{equation}

This yields the following expressions for the "weighted” order parameter:
\begin{subequations}
\label{A_wL}
\begin{align}
\mathcal{W}_1 &=
\frac{\Gamma\!\left(\tfrac{d}{2}\right)}{\det \mathcal{W}}
\Bigl[
\Gamma\!\left(1-\tfrac{d}{2}\right)
\cos\!\Bigl(\tfrac{\pi(d-1)}{2}\Bigr)
L^{\frac{d}{2}-1}
e^{\mathrm{i}\varphi_\varepsilon}
\nonumber\\[-2pt]
&\qquad\quad
+\,\pi\!\left(\tfrac{\lambda}{2}\right)^{1-\frac{d}{2}}
Y_{\frac{d}{2}-1}(\lambda L)
\Bigr],
\\[4pt] 
\mathcal{W}_2 &=
\frac{\pi}{\det \mathcal{W}}
\Bigl[
\Gamma\!\left(\tfrac{d}{2}\right)
\!\left(\tfrac{\lambda}{2}\right)^{1-\frac{d}{2}}
J_{\frac{d}{2}-1}(\lambda L)
-
L^{\frac{d}{2}-1}
e^{\mathrm{i}\varphi_\varepsilon}
\Bigr],
\end{align}
\end{subequations}
where we define the new phase $\varphi_\varepsilon \equiv \varphi - \varepsilon L/2\gamma$ and $\det \mathcal{W} = \Gamma \left( {\frac{d}{2}} \right)\Gamma \left( {1 - \frac{d}{2}} \right)\cos \left( {\frac{{\pi \left( {d - 1} \right)}}{2}} \right){J_{\frac{d}{2} - 1}}\left( {\lambda L} \right) + \pi {Y_{\frac{d}{2} - 1}}\left( {\lambda L} \right)$.

For the ”unweighted” counterpart we have:
\begin{subequations}
\label{A_uL}
\begin{align}
\mathcal{U}_1 &=
\frac{1}{\det \mathcal{U}}
\Bigl[
\frac{\Gamma(2-d)}{\Gamma\!\left(\tfrac{3-d}{2}-\kappa\right)}
L^{\frac{d-1}{2}}
e^{i\varphi_\varepsilon}
\nonumber\\[-2pt]
&\qquad\quad
-\,(2i\lambda)^{\frac{1-d}{2}}
W_{\kappa,\frac{d}{2}-1}(2i\lambda L)
\Bigr],
 \\[4pt]
\mathcal{U}_2 &=
\frac{1}{\det \mathcal{U}}
\Bigl[
(2i\lambda)^{\frac{1-d}{2}}
M_{\kappa,\frac{d}{2}-1}(2i\lambda L)
-
L^{\frac{d-1}{2}}
e^{i\varphi_\varepsilon}
\Bigr],
\end{align}
\end{subequations}
where another determinant is introduced $\det \mathcal{U} = \frac{{\Gamma \left( {2 - d} \right)}}{{\Gamma \left( {\frac{{3 - d}}{2} - \kappa } \right)}}{M_{\kappa ,\frac{d}{2} - 1}}(2i\lambda L) - {W_{\kappa ,\frac{d}{2} - 1}}(2i\lambda L)$.

\begin{figure*}
\centering
\includegraphics[width=0.5\linewidth]{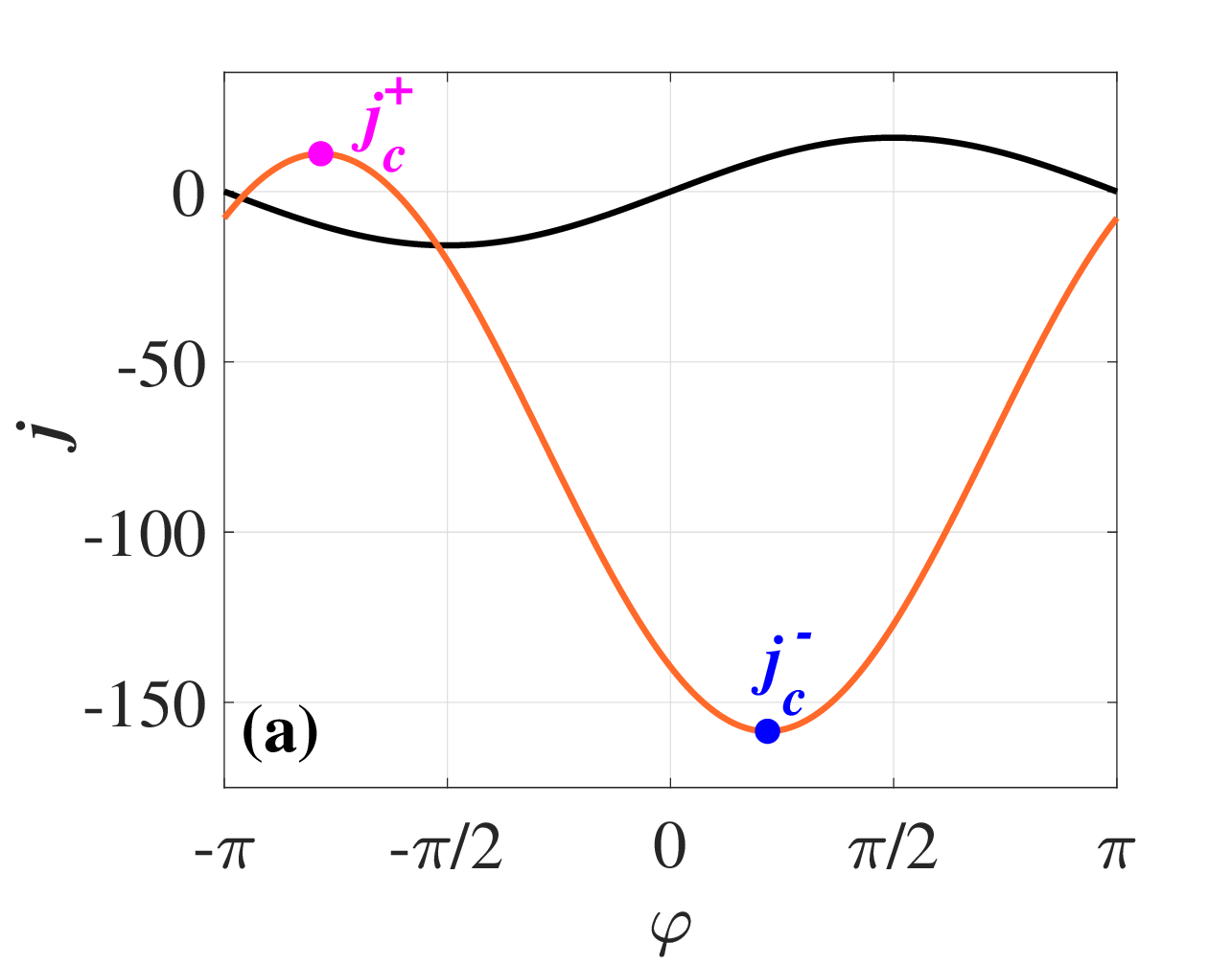}\hfil
\includegraphics[width=0.5\linewidth]{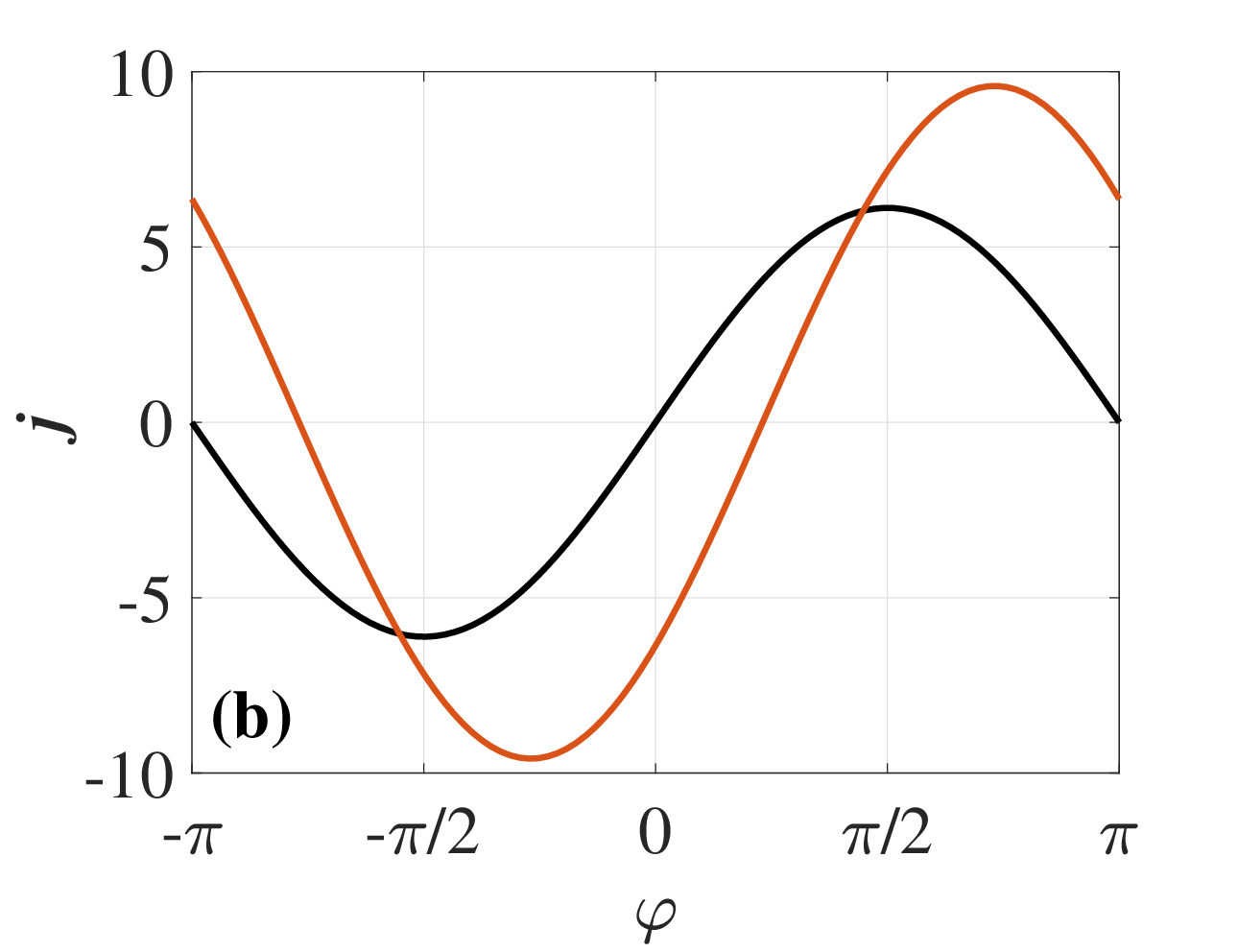}
\caption{Current-phase relations for the Josephson junction with the length $L=5$ between fractal superconductors with dimensionality $d=0.9$ calculated within "unweighted" (a) and "weighted" (b) GL formalism in the absence of non-Hermitian Lifshitz drift when $\varepsilon=0$ (black lines) and in its presence when $\varepsilon/\sqrt{|a|\gamma} \approx 1.547$ (orange lines). In Figure (a), we additionally label the positions of the positive and negative critical currents to accent  the emergence of a giant diode effect with $\eta = -0.869$. The  amplitude of the supercurrent density is taken in units of $2e\gamma |\Delta_0|^2$.}
\label{CPR_frac_int}
\end{figure*}

In the limit $d=1$, using the relations of Whittaker and Bessel functions to hyperbolic functions when $\nu=-1/2$ \cite{abramowitz}  one can simplify both expressions for the order parameter Eqs. \eqref{OP_wL} and Eqs. \eqref{OP_uL} to a superposition of two exponents, an ansatz that was applied for a quasi-one-dimensional Josephson junction between conventional superconductors, when the normal layer is a noncentrosymmetric metal \cite{Buzdin}.

\begin{figure*}
\centering
\includegraphics[width=0.49\linewidth]{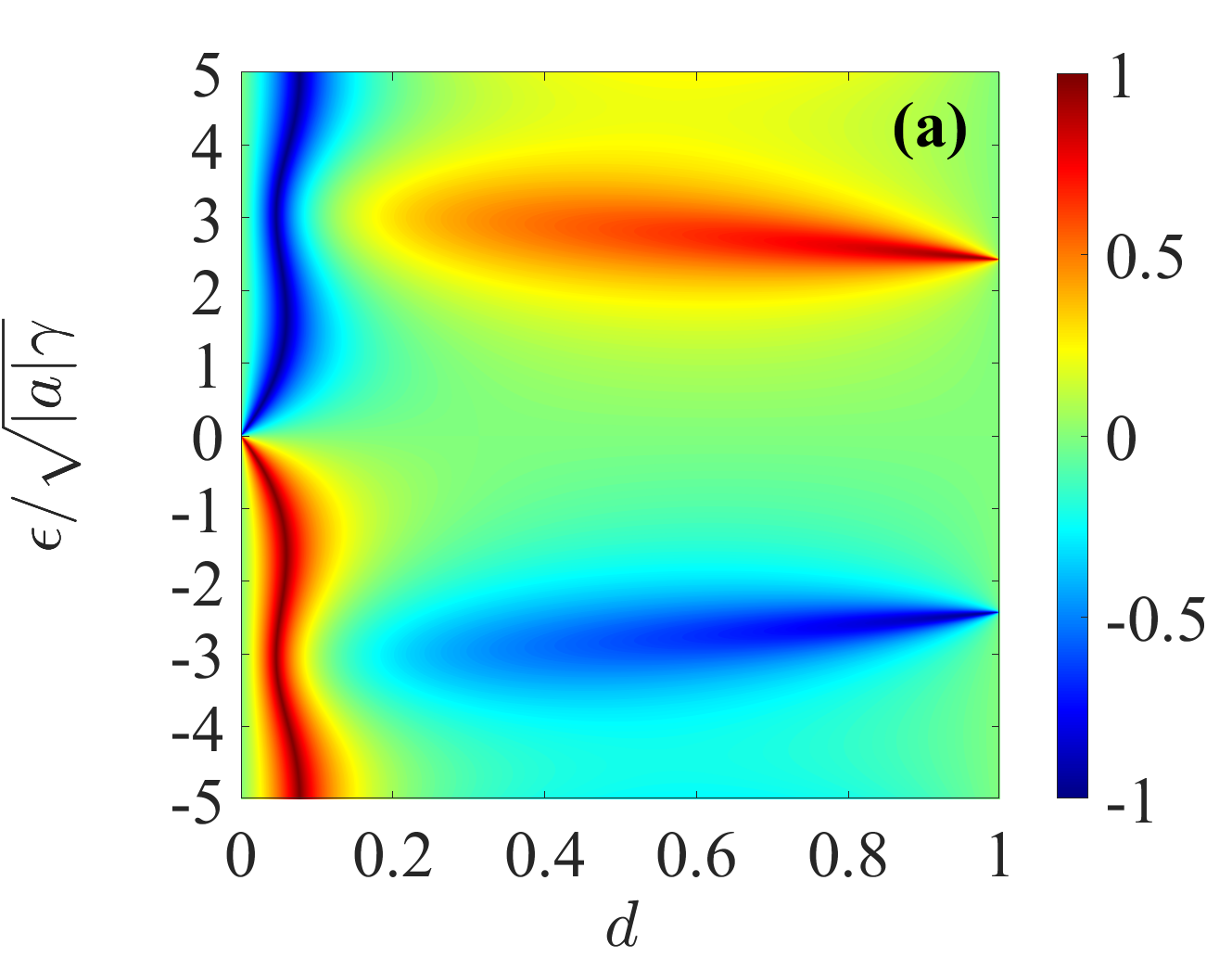}\hfil
\includegraphics[width=0.49\linewidth]{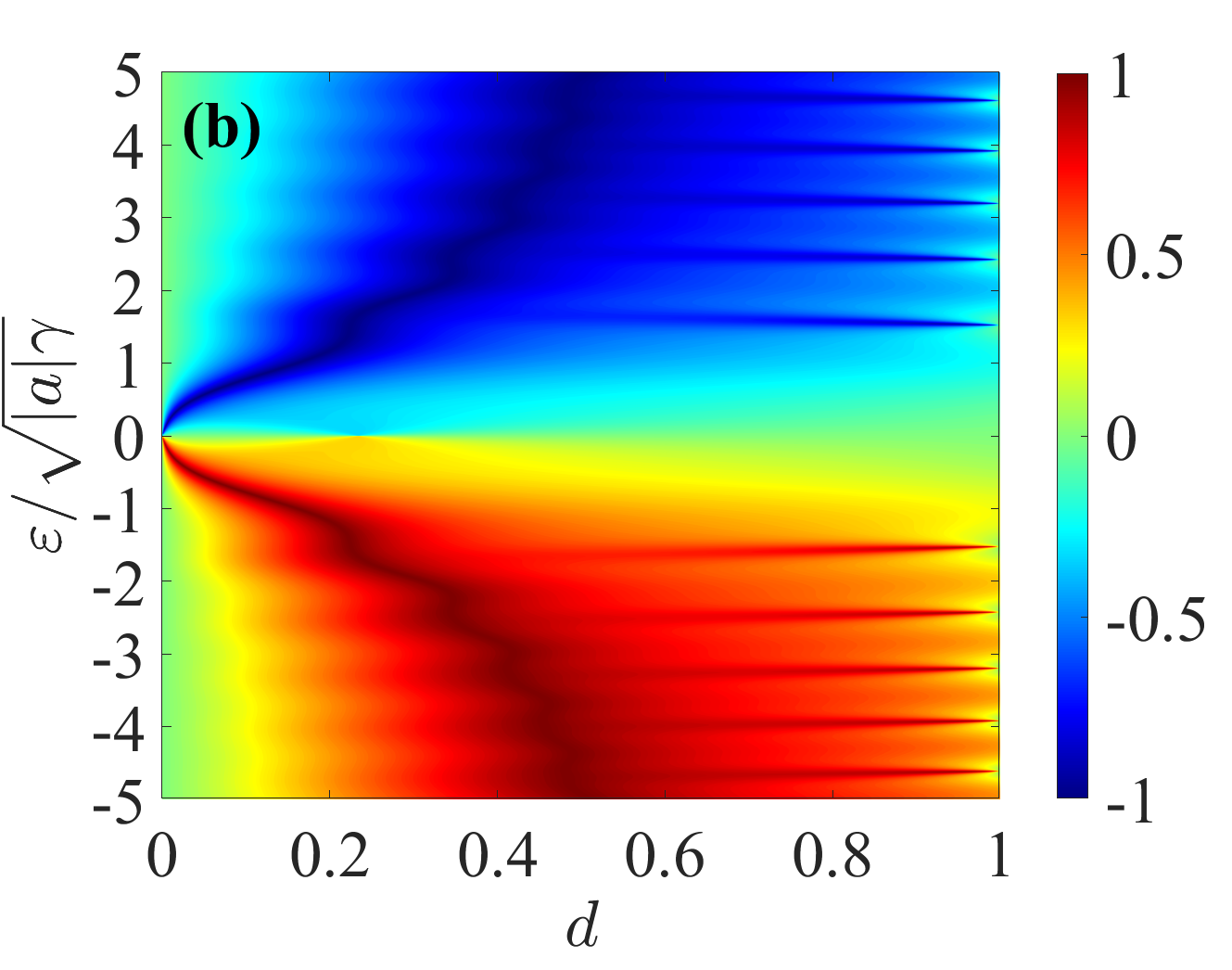}\par\medskip
\includegraphics[width=0.49\linewidth]{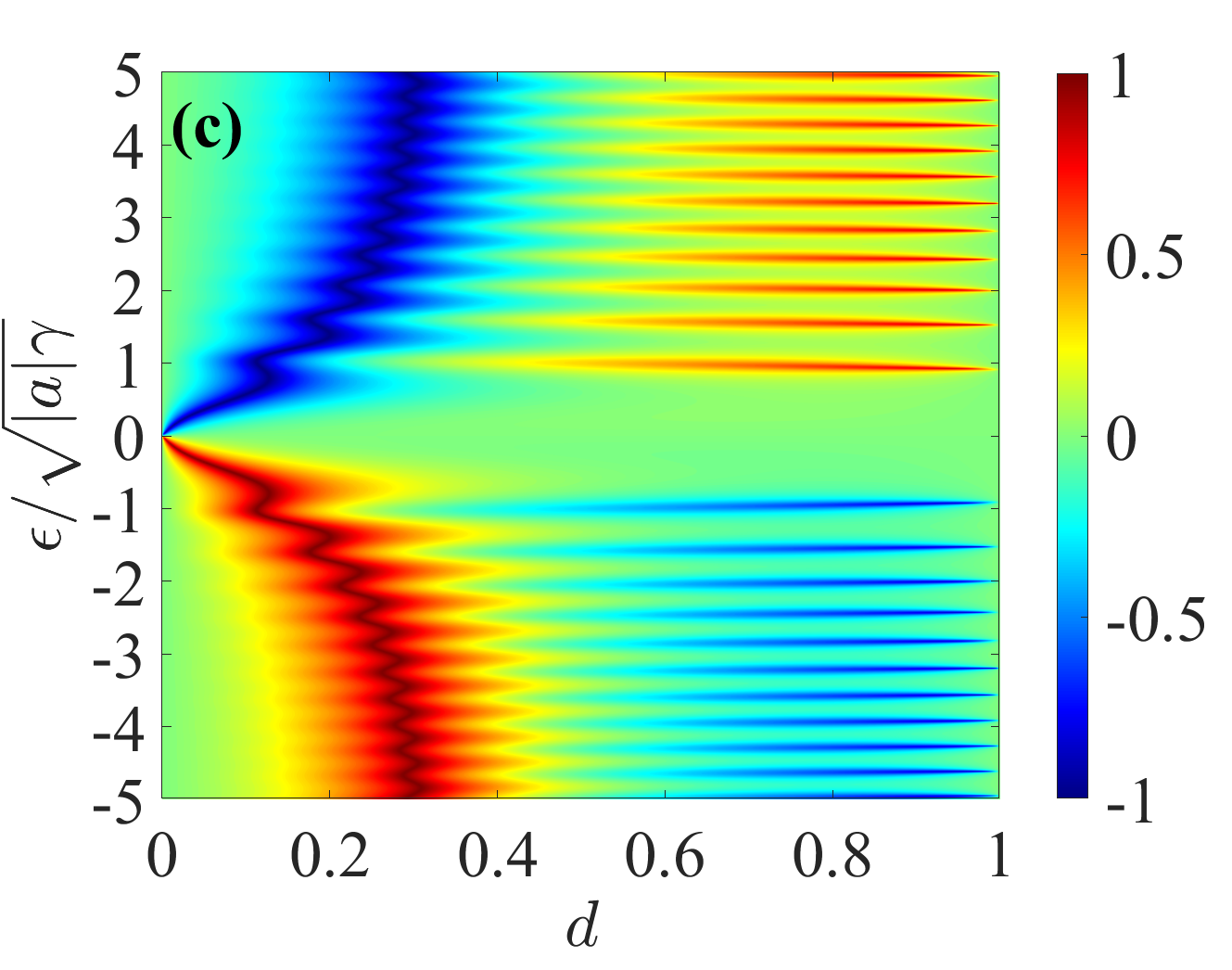}\hfil
\includegraphics[width=0.49\linewidth]{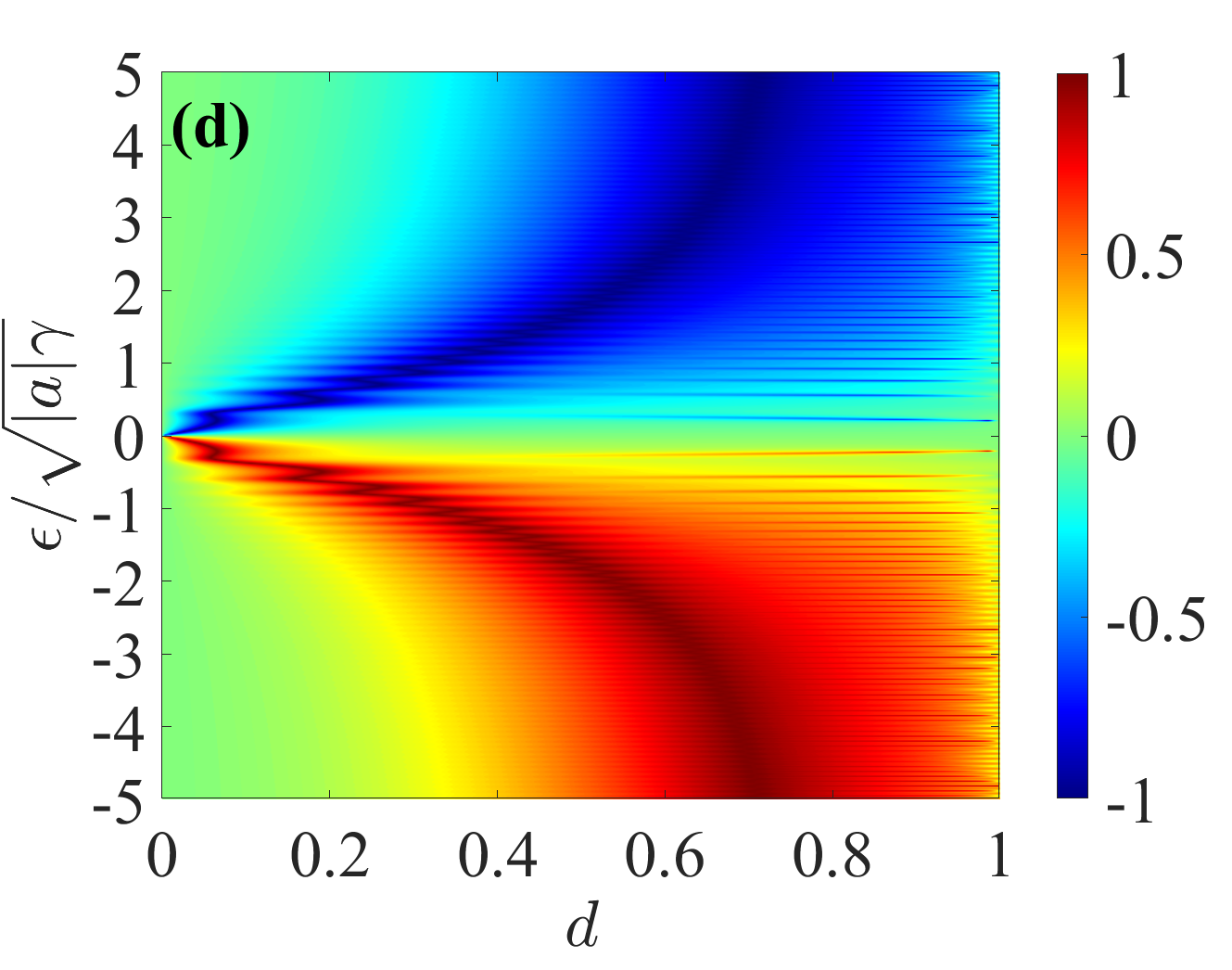}
\caption{Contour map of the supercurrent rectification amplitude $\eta$ for the Josephson junction as a function of the  fractional dimensionality $0<d \leq 1$ and the coefficient $\epsilon/\sqrt{|a|\gamma}$ for the length $L=\sqrt{\gamma/\|a|}$ (a), $L=5\sqrt{\gamma/\|a|}$ (b), $L=10\sqrt{\gamma/\|a|}$ (c) and $L=50\sqrt{\gamma/\|a|}$ (d) calculated within "unweighted" GL formalism.}
\label{diode_eff_Whittaker}
\end{figure*}

Since the central focus of our research is the identification of the nonreciprocal transport properties of the Josephson system, we derive an expression for the current density based on the GL functional Eq. (\ref{GL_free_energy}). It can be written as follows:
\begin{align}
j(x)=
e\,\frac{|x|^{d-1}}{\Gamma(d)}
\Bigl[
i\gamma
\bigl(\Delta(x)\Delta^{*'}(x)-\Delta^{*}(x)\Delta'(x)\bigr)
\nonumber \\[-2pt]
\qquad
-\,\varepsilon\,w(x)\,|\Delta(x)|^{2}
\Bigr].
\label{current_density}
\end{align}

Using Eq. (\ref{current_density}) and taking into account the expressions for the order parameter Eqs. (\ref{OP_wL}) and (\ref{OP_uL}) with coefficients given by Eqs. (\ref{A_wL}) and (\ref{A_uL}), we can obtain the current-phase relations for a Josephson junction. We do not give them here explicitly because of the very cumbersome expressions, which are very difficult to examine analytically. For this reason, numerical analysis is applied, where the diode rectification amplitude, $\eta$, is determined by the difference between the maximal amplitudes of the critical currents for forward $I_c^+$ and backward $I_c^-$ flow directions:
\begin{equation}
\eta=\frac{j_c^{(+)}-|j_c^{(-)}|}{j_c^{(+)}+|j_c^{(-)}|}.
\label{diode_gen}
\end{equation}

Figure \ref{CPR_frac_int} displays the current-phase relations of a Josephson junction with length $L=5$ in units $\sqrt{\gamma/\|a|}$ calculated within the "unweighted" and "weighted" approach of fractional GL theory. As can be seen in Figure \ref{CPR_frac_int}a, the interplay of the Lifshitz invariant and fractional dimension generates nonreciprocal transport, which are two crucial ingredients for the emergence of the diode effect between fractal superconductors. Moreover, it turns out that nonreciprocal transport and diode effect are realized only in the case of the “unweighted” formulation. The origin of this behavior stems from the mathematical structure of the solutions for the order parameter.

In the weighted formulation, the linearized GL operator underlying Eq.~(\ref{OP_wL}) becomes real and self-adjoint. Together with symmetric Josephson boundary conditions, the boundary-value problem is invariant under reversing the coordinate across the link (\(x\!\to\!L-x\)), which enforces an odd current-phase relation \(I(\varphi)=-I(-\varphi)\) and hence equal forward and backward critical currents, \(I_c^{(+)}=I_c^{(-)}\) (no diode effect). By contrast, in the unweighted formulation, the reduced equation is non-self-adjoint and maps to a Whittaker solution Eq.~(\ref{OP_uL}) with a nonzero parameter $\kappa$ that encodes a directional bias of the order-parameter profile. This non-removable bias generates even-in-\(\varphi\) contribution to the current, so \(I_c^{(+)}\neq I_c^{(-)}\) and a finite diode efficiency emerges.

The Fourier analysis performed reveals (see Appendix \ref{App_A}) that the current-phase relations in the latter case can be characterized by the general formula:
\begin{equation}
j(\varphi)=\mathcal{A}\sin{\varphi}+\mathcal{B}\cos{\varphi}+\mathcal{C},
\label{CPR_unweighted}
\end{equation}
with the above-mentioned even-in-\(\varphi\) cosine contribution, where $\mathcal{A}$, $\mathcal{B}$ and $\mathcal{C}$ are harmonic amplitudes. 

Finally, as numerical calculations show, for certain values of the coefficient $\varepsilon$ and the fractional dimension of the space $d$, it is possible to achieve even a practically perfect diode effect, i.e., the rectification amplitude coefficient is close to $\eta \approx \pm 1$ (see Fig. \ref{diode_eff_Whittaker}). A remarkable feature is that as the length $L$ of the Josephson junction increases, the diode efficiency pattern becomes highly oscillatory (see Fig. \ref{diode_eff_Whittaker}b-d). At the same time, regardless of $L$, one of the domains where $\eta$ reaches an almost perfect efficiency is the region where the dimension $d$ is close to $1$ with $\varepsilon \ne 0$ but nevertheless not equal to it.

\section{Fractional generalization of the GL equation}

To generalize the GL equation for the fractional case, we use the Caputo fractional derivative \cite{Caputo}. They are the suitable tool for the Josephson weak link under consideration because they respect the physics at the boundaries and the gauge: they work directly with Dirichlet boundary conditions at $x=0$ and $x=L$, and they give zero kinetic response to a uniform phase, so the definition of a gauge-invariant current can be clean. They also admit a tidy variational formulation in which the left and right sides are adjoints, letting us derive the GL equation and the current density from the corresponding functional.

Crucially for diode physics, Caputo’s one-sided kernels let us weight left and right memory differently, cleanly encoding the structural asymmetry of Josephson contacts driving nonreciprocity. At the same time, the operators reduce smoothly to the usual first/second derivatives in the local limit, so the fractional model nests the classical integer-order GL theory. Practically, they offer analytic tractability and good numerical implementation, avoiding possible endpoint singularities. By contrast, Riemann–Liouville fractional derivatives often require fractional initial data and react spuriously to constants, while symmetric Riesz operators are excellent for periodic or bulk media but lack directional control at finite boundaries—precisely what our asymmetric Josephson junction needs.

It is important to note that we do not claim a one-to-one mapping between any specific fractal geometry of superconductors and a unique fractional order. Instead, we apply fractional operators as an effective continuum representation that captures the anomalous transport observed across the Josephson junction.

For the interval $x\in[0,L]$ one can introduce the left- and right-side Caputo fractional derivatives of order $n-1<\nu\le n$ with $n\in\{1,2\}$ for a continuous function $f(x)$ \cite{Caputo}:
\begin{equation}
\begin{aligned}
{}_{0}^{C}D_{x}^{\nu} f(x) 
&= \frac{1}{\Gamma(n-\nu)} \int_{0}^{x} 
   \frac{f^{(n)}(\xi)}{(x-\xi)^{\nu-n+1}} \, d\xi, \\[8pt]
{}_{x}^{C}D_{L}^{\nu} f(x) 
&= \frac{(-1)^{n}}{\Gamma(n-\nu)} \int_{x}^{L} 
   \frac{f^{(n)}(\xi)}{(\xi-x)^{\nu-n+1}} \, d\xi,
\end{aligned}
\end{equation}
where, for convenience, we denote this type of fractional derivatives by means of the superscript $C$.

The Agrawal variational principle for fractional derivatives \cite{Agrawal}, which extends the classical Euler-Lagrange equation, allows us to generalize the linearized version of the GL equation for fractional orders:
\begin{widetext}
\begin{equation}
-\gamma \left[ \kappa \, {}_0^C D_x^\beta \Delta(x) + (1 - \kappa) \, {}_x^C D_L^\beta \Delta(x) \right] + i \varepsilon \left[ \mu \, {}_0^C D_x^\sigma \Delta(x) + (1 - \mu) \, {}_x^C D_L^\sigma \Delta(x) \right] + \alpha \Delta(x) = 0,
\label{eq:fractional_gl}
\end{equation}
\end{widetext}
where $0<\sigma < 1$, $1<\beta < 2$ are fractional orders of derivatives for Lifshitz and kinetic terms, respectively, and $0\le\kappa,\mu\le 1$ determine corresponding left and right partial contributions (mixing weights) of these terms. We emphasize that the fractional orders ( $\sigma$, $\beta$) are effective transport parameters inferred from anomalous dispersion/response; no unique mapping to a specific fractal is implied.

Based on the Agrawal variational principle, it is possible to restore the GL free energy, to re-derive Eq. \eqref{eq:fractional_gl} and then obtain the expression for the current density (see Appendix \ref{App_B} for details):
\begin{equation}
j(x)= j_{\beta}(x)+j_{\sigma}(x),
\end{equation}
where $j_{\beta}(x)$ denotes the kinetic contribution
%\begin{equation}
%j_{\beta}(x) = 2e\gamma\,\Im\!\Big\{
%\Delta^*(x)\,\Big[
%\kappa\,{}_{0}I_{x}^{\,2-\beta}\,{}_{0}^{C}\!D_{x}^{\beta}
%-(1-\kappa)\,{}_{x}I_{L}^{\,2-\beta}\,{}_{x}^{C}\!D_{L}^{\beta}
%\Big]\Delta(x)\Big\},
%\label{j_beta}
%\end{equation}
\begin{equation}
\label{j_beta}
\begin{aligned}
j_{\beta}(x)
&= 2e\gamma\,\Im\!\Big\{
\Delta^*(x)\,
\Big[
\kappa\,{}_{0}I_{x}^{\,2-\beta}\,{}_{0}^{C}\!D_{x}^{\beta}
\\
&\qquad\qquad
-(1-\kappa)\,{}_{x}I_{L}^{\,2-\beta}\,{}_{x}^{C}\!D_{L}^{\beta}
\Big]\,
\Delta(x)
\Big\}.
\end{aligned}
\end{equation}
and $j_{\sigma}(x)$ corresponds to the non-Hermitian (Lifshitz) part
%\begin{equation}
%j_{\sigma}(x) = -e\varepsilon\,\Re\!\Big\{
%\Delta^*(x)\,\Big[
%\mu\,{}_{0}I_{x}^{\,1-\sigma}\,{}_{0}^{C}\!D_{x}^{\sigma}
%-(1-\mu)\,{}_{x}I_{L}^{\,1-\sigma}\,{}_{x}^{C}\!D_{L}^{\sigma}
%\Big]\Delta(x)\Big\}.
%\label{j_sigma}
%\end{equation}
\begin{equation}
\label{j_sigma}
\begin{aligned}
j_{\sigma}(x)
&= -e\varepsilon\,\Re\!\Big\{
\Delta^*(x)\,
\Big[
\mu\,{}_{0}I_{x}^{\,1-\sigma}\,{}_{0}^{C}\!D_{x}^{\sigma}
\\
&\qquad\qquad
-(1-\mu)\,{}_{x}I_{L}^{\,1-\sigma}\,{}_{x}^{C}\!D_{L}^{\sigma}
\Big]\,
\Delta(x)
\Big\}.
\end{aligned}
\end{equation}

Here $_{0}I_{x}^{\,\rho}$ and $_{x}I_{L}^{\,\rho}$ are left- and right-sided Riemann–Liouville fractional integrals of order $\rho$. They are structurally required by the fractional variational calculus to get a well-posed GL functional, a proper current density, and correct boundary behavior in Agrawal fractional variational calculus.

\subsubsection{Plane-wave solution}

Using identities ${}_{0}^{C}\!D_{x}^{\nu} e^{q x}=q^{\nu}e^{q x}$, ${}_{x}^{C}\!D_{L}^{\nu} e^{q x}=(-q)^{\nu}e^{q x}$, with principal branches for the fractional powers, one can apply a plane-wave ansatz $\Delta(x)=e^{q x}$ and substitute it Eq. \eqref{eq:fractional_gl}. This yields the characteristic dispersion relation:
%\begin{equation}
%-\gamma\!\left[\kappa\,q^{\beta}+(1-\kappa)(-q)^{\beta}\right]
%+i\varepsilon\!\left[\mu\,q^{\sigma}+(1-\mu)(-q)^{\sigma}\right]
%+\alpha \;=\; 0.
%\label{eq:char}
%\end{equation}
\begin{equation}
\label{eq:char}
\begin{gathered}
-\gamma\!\left[\kappa\,q^{\beta}+(1-\kappa)(-q)^{\beta}\right]
\\
+\;i\varepsilon\!\left[\mu\,q^{\sigma}
+(1-\mu)(-q)^{\sigma}\right] + a
= 0 .
\end{gathered}
\end{equation}

The fractional nature of our approach does not invalidate this, as the Caputo fractional derivative of an exponential yields a scaled exponential, maintaining linearity. Denoting  $q_j$ as any two distinct roots $q_1\neq q_2$ of Eq. \eqref{eq:char} one can represent the solution of Eq. \eqref{eq:fractional_gl} in the form:
\begin{equation}
\Delta(x)=C_1 e^{q_1 x}+C_2 e^{q_2 x},
\label{eq:gensol}
\end{equation}
where the constants $C_1,C_2$ can be found by means of the Dirichlet boundary conditions Eq. \eqref{eq:BC}:
\begin{equation}
\label{eq:C12}
\begin{aligned}
C_1=\frac{|\Delta_0|\,(e^{i\varphi}-e^{q_2 L})}{e^{q_1 L}-e^{q_2 L}}, \\
C_2=\frac{|\Delta_0|\,(e^{q_1 L}-e^{i\varphi})}{e^{q_1 L}-e^{q_2 L}}.
\end{aligned}
\end{equation}

As a result, the final expression for the order parameter is given by:
\begin{equation}
\Delta(x)\!=\!|\Delta_0|\,
\frac{(e^{i\varphi}\!-\!e^{q_2 L})\,e^{q_1 x}\!+\!(e^{q_1 L}\!-\!e^{i\varphi})\,e^{q_2 x}}
{e^{q_1 L}\!-\!e^{q_2 L}}\,,
\label{OP_plane_wave}
\end{equation}
From the physical point of view it is necessary to select a pair of wave-vectors with $\Re q_{1,2}$ of opposite sign to control growth and decay across the Josephson junction.

For $\beta = 2$, $\sigma = 1$, we can recover the  quadratic dispersion relation $-\gamma q^2+i\varepsilon(2\mu-1)q+a=0$, which gives the same set of wave-vectors $q_{1,2}$ as in Ref. \onlinecite{Buzdin} (see Appendix \ref{App_C} for consistency check).

For $1<\beta<2$ the Caputo fractional derivatives can be written as Riemann–Liouville fractional integrals of the second derivative, and then using their semigroup property we obtain:
\begin{equation}
\label{eq:undo-beta_main}
\begin{aligned}
{}_0 I_x^{\,2-\beta}\,{}_0^C D_x^\beta \Delta=\Delta'(x)-\Delta'(0), \\
{}_x I_L^{\,2-\beta}\,{}_x^C D_L^\beta \Delta=-\Delta'(x)+\Delta'(L).
\end{aligned}
\end{equation}

Substituting Eq. \eqref{eq:undo-beta_main} into the current density component Eq. \eqref{j_beta}  and as in the previous section, evaluating at $x=L$ we obtain:
\begin{equation}
j_\beta(\varphi) =J_{\beta}(q_1,q_2,L)\,\sin\varphi,
\label{eq:jbeta-odd}
\end{equation}
with the prefactor
%\begin{equation}
%J_{\beta} = \frac{{2e\gamma {\mkern 1mu} |{\Delta _0}{|^2}}}{{{{\left| {{e^{{q_1}L}} - {e^{{q_2}L}}} \right|}^2}}} {\mathop{\rm Re}\nolimits} \left\{ {{{\left( {{e^{{q_1}L}} - {e^{{q_2}L}}} \right)}^ * }\left[ {{q_1}\left( {{e^{{q_1}L}} - 1} \right) - {q_2}\left( {{e^{{q_2}L}} - 1} \right)} \right]} \right\},
%\label{eq:J1beta}
%\end{equation}
\begin{equation}
\label{eq:J1beta}
\begin{split}
J_{\beta} &=
\frac{2e\gamma\,|\Delta_0|^2}
     {\bigl|e^{q_1 L}-e^{q_2 L}\bigr|^2}
\\
&\quad\times
\Re\!\left\{
\bigl(e^{q_1 L}\!-\!e^{q_2 L}\bigr)^{\!*}
\!\Bigl[
q_1\bigl(e^{q_1 L}\!-\!1\bigr)
\!- \!q_2\bigl(e^{q_2 L}\!-\!1\bigr)
\Bigr]
\right\}.
\end{split}
\end{equation}

Although $\kappa$ and $\mu$ do not appear explicitly in \eqref{eq:J1beta}, they enter implicitly through the roots $q_{1,2}$ of the $\kappa$- and $\mu$-dependent characteristic dispersion Eq. \eqref{eq:char}.

For $0<\sigma < 1$ the Caputo-Riemann–Liouville identities are
\begin{equation}
\label{eq:undo-sigma}
\begin{aligned}
{}_0 I_x^{\,1-\sigma}{}_0^C D_x^\sigma \Delta=\Delta(x)-\Delta(0),\\
{}_x I_L^{\,1-\sigma}{}_x^C D_L^\sigma \Delta=\Delta(x)-\Delta(L).
\end{aligned}
\end{equation}

A direct evaluation yields the $\sigma$-contribution term of the current density:
\begin{equation}
j_{\sigma}(\varphi)=-e\varepsilon\,|\Delta_0|^2\,(1-\cos\varphi)\,,
\label{eq:even-unique}
\end{equation}

Collecting Eqs. \eqref{eq:jbeta-odd} and \eqref{eq:even-unique}, the current-phase relation acquires the final form:
\begin{equation}
j(\varphi)=J_{\beta}\,\sin\varphi
-e\varepsilon\,|\Delta_0|^2\,(1-\cos\varphi)\,,
\label{eq:CPR-final}
\end{equation}
where $J_{\beta}$ is given in Eq. \eqref{eq:J1beta}. 

\begin{figure*}
\centering
\includegraphics[width=0.49\linewidth]{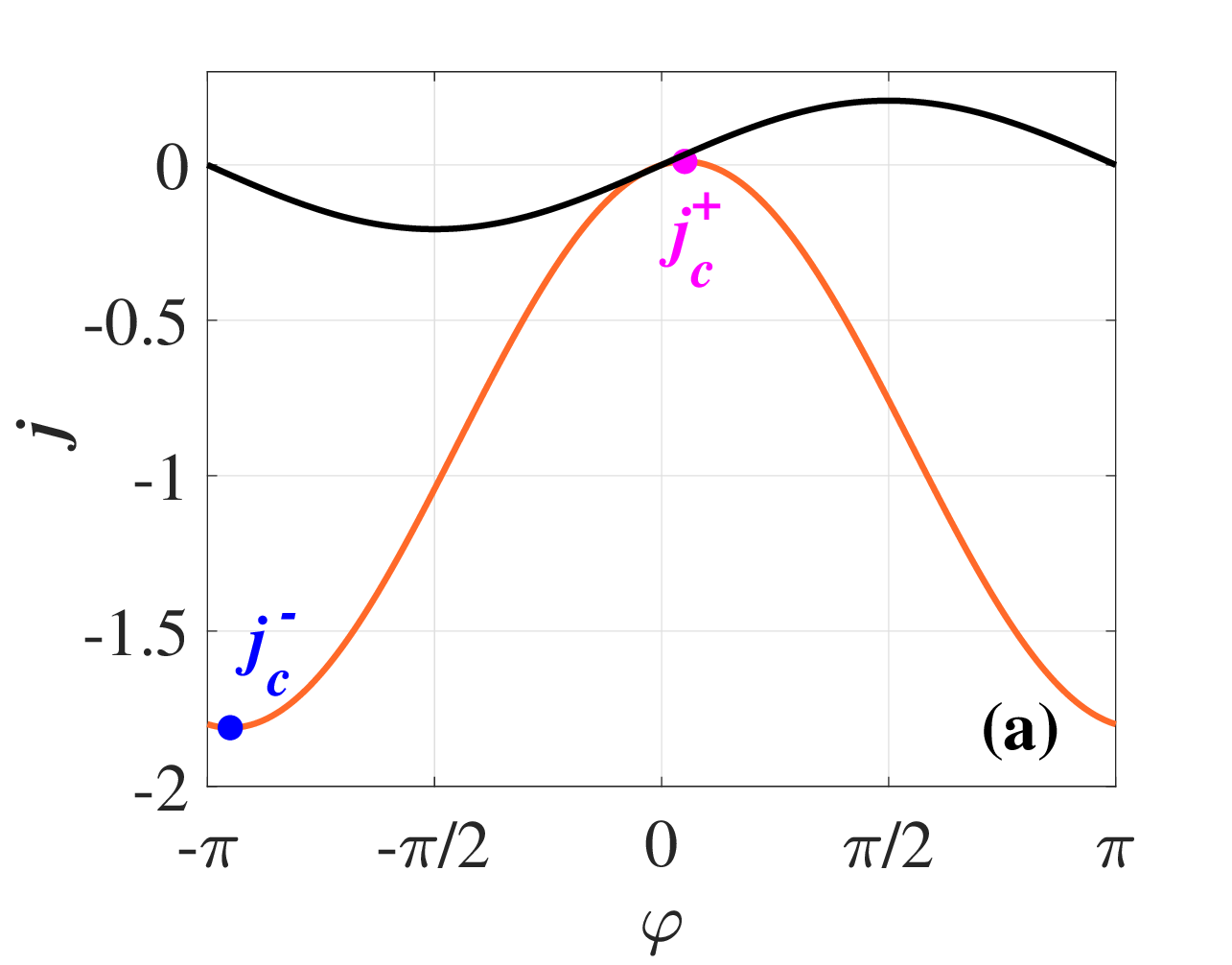}\hfil
\includegraphics[width=0.49\linewidth]{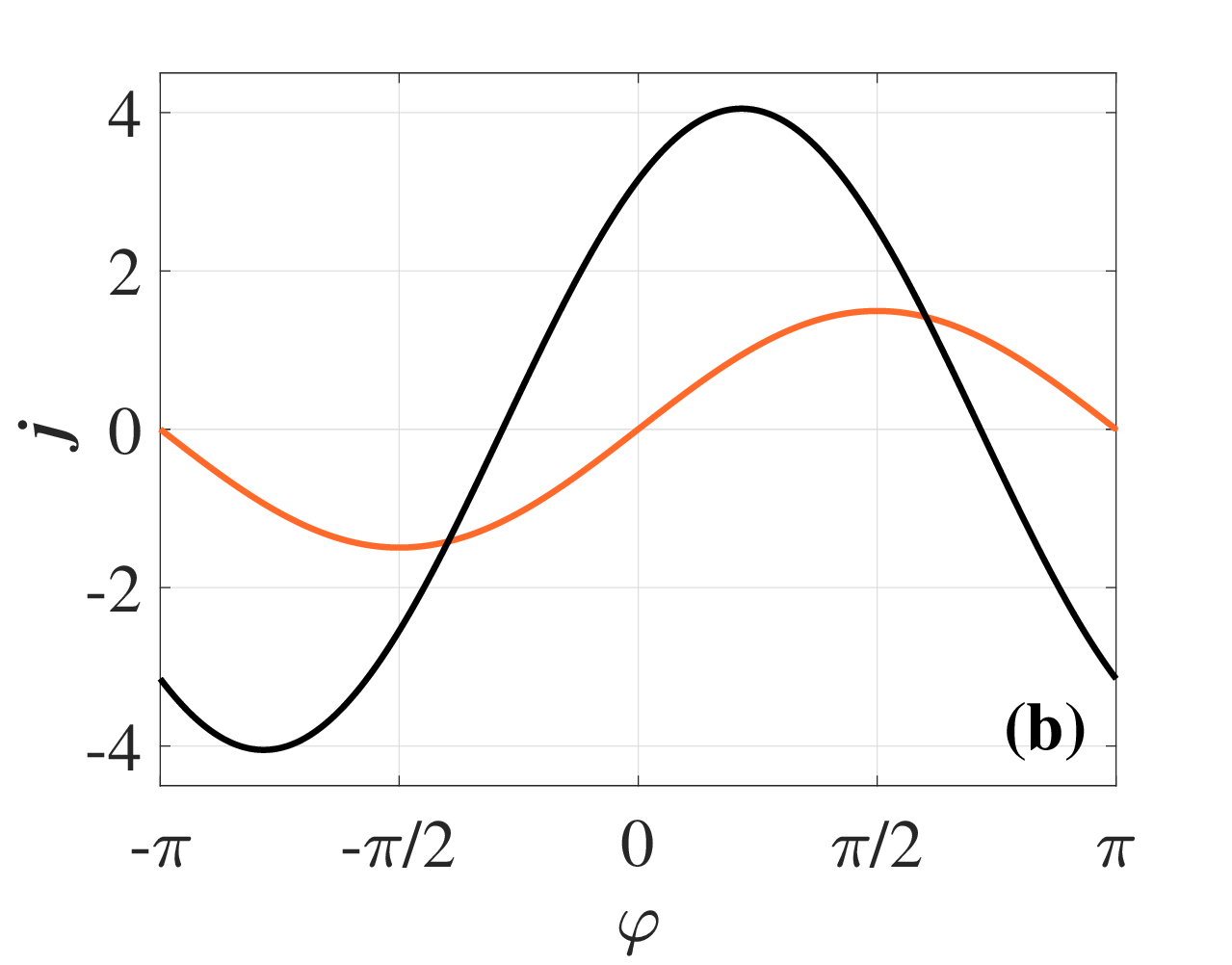}\par\medskip
\caption{(a) Current-phase relations for the Josephson junction with the length $L=5$ between fractal superconductors calculated for fractional orders of the Caputo derivatives $\beta=1.5$, $\sigma=0.9$  in the absence of non-Hermitian Lifshitz drift when $\varepsilon=0$ (black line) and in its presence when $\varepsilon/\sqrt{|a|\gamma} =0.9$ (orange line) for equal left- and right-sided contributions $\kappa=\mu=0.5$. (b) Current-phase relations for the Josephson junction of the same length $L$ calculated for integer orders of the Caputo derivatives $\beta=2$, $\sigma=1$ for the fixed value of $\varepsilon/\sqrt{|a|\gamma} =0.9$ with the pure right-sided kinetic mixing weight $\kappa=0$ and pure left-sided Lifshitz mixing weight $\mu=1$ (black line) and with their equal contributions $\kappa=\mu=0.5$ (orange line).
The  amplitude of the supercurrent density is taken in units of $2e\gamma |\Delta_0|^2$.}
\label{CPR}
\end{figure*}

The visualization of the current-phase relations described by Eq. \eqref{eq:CPR-final} is shown in Figure \ref{CPR}. It consists of two plots. The orange curve in Figure \ref{CPR}a illustrates the strong nonreciprocal response ($\eta \approx -1$) of the Josephson junction emerging from the interplay of the Lifshitz term with the fractional kinetics, represented by fractional Caputo derivatives of the order of $\beta$ and $\sigma$. In turn, the black curve in the same plot emphasizes the importance of the Lifshitz invariant, when for $\varepsilon=0$ the current-phase relation becomes reciprocal regardless of fractional values of $\beta$ and $\sigma$. 

The second part of Figure \ref{CPR}, i.e., Figure \ref{CPR}b is devoted to the current-phase relations calculated in the integer limit of Caputo fractional derivatives $\beta=2$ and $\sigma=1$ with their left-sided and right-sided contributions $\kappa$ and $\mu$ to the kinetic and Lifshitz terms, respectively, in the GL Eq. (\ref{eq:fractional_gl}).  It can be seen that for $\kappa=0$ and $\mu=1$, the reciprocal current-phase relation of the Josephson $\varphi_0$ weak link is reproduced (black curve). It is noteworthy that in the case of equal left- and right-sided contributions $\kappa=\mu=0.5$, the Josephson weak link acquires a classical sinusoidal profile of the dependence $j(\varphi)$ (orange curve). The reason for the latter result stems from the fact that the coefficient $\varepsilon$ of the Lifshitz invariant for $\mu=0.5$ effectively becomes equal to $\varepsilon_{\rm eff} \;=\; (2\mu-1)\,\varepsilon=0$ (see Appendix \ref{App_C}).

At first glance, from Eq. \eqref{eq:CPR-final} follows an unusual result for $\varphi=0$ we have $j=0$, which contradicts the previously obtained current-phase relation Eq. \eqref{CPR_unweighted}, when $\varphi=0$ yields nonzero current density. However, this does not compromise the fidelity of the current-phase relation for $\varphi\neq 0$. First, the odd Josephson amplitude $J_{\beta}$ and the local slope are captured exactly $j'(0)=J_{\beta}$, which coincides with the first harmonic of the full solution. Second, derived within the plane-wave anstz current-phase relation can be rewritten as $j(\varphi)=R\sin(\varphi+\varphi_{0})-P$ with
$R=\sqrt{J_{\beta}^{2}+P^{2}}$ and $\varphi_0=\arctan(P/J_{\beta})$. Hence, the position of the extrema and their nonreciprocal splitting are determined by $R$ and $B$ and are independent of zeroing at $\varphi=0$. In other words, for $I_{c}^{(+)}=R-P$, $I_{c}^{(-)}=R+P$, so $I_{c}^{(+)}\neq I_{c}^{(-)}$ for any $B\neq 0$. In short, the zero current density at $\varphi=0$ is a point artifact, while the $\varphi$-dependent features that control nonreciprocity (slope, extrema values, and their asymmetry) remain accurately taken into account by the two–mode plane wave approximation.

Therefore, within the Caputo–based fractional GL framework, i.e. Agrawal variational calculus with left/right Caputo kernels, the gauge–invariant transport current produces the current–phase relation Eq. \eqref{eq:CPR-final}, where the  prefactor $J_{\beta}$ encodes the fractional kinetics through the dispersion roots $q_{1,2}$ and implicitly the mixing weights. 

\begin{figure*}
\centering
\includegraphics[width=0.49\linewidth]{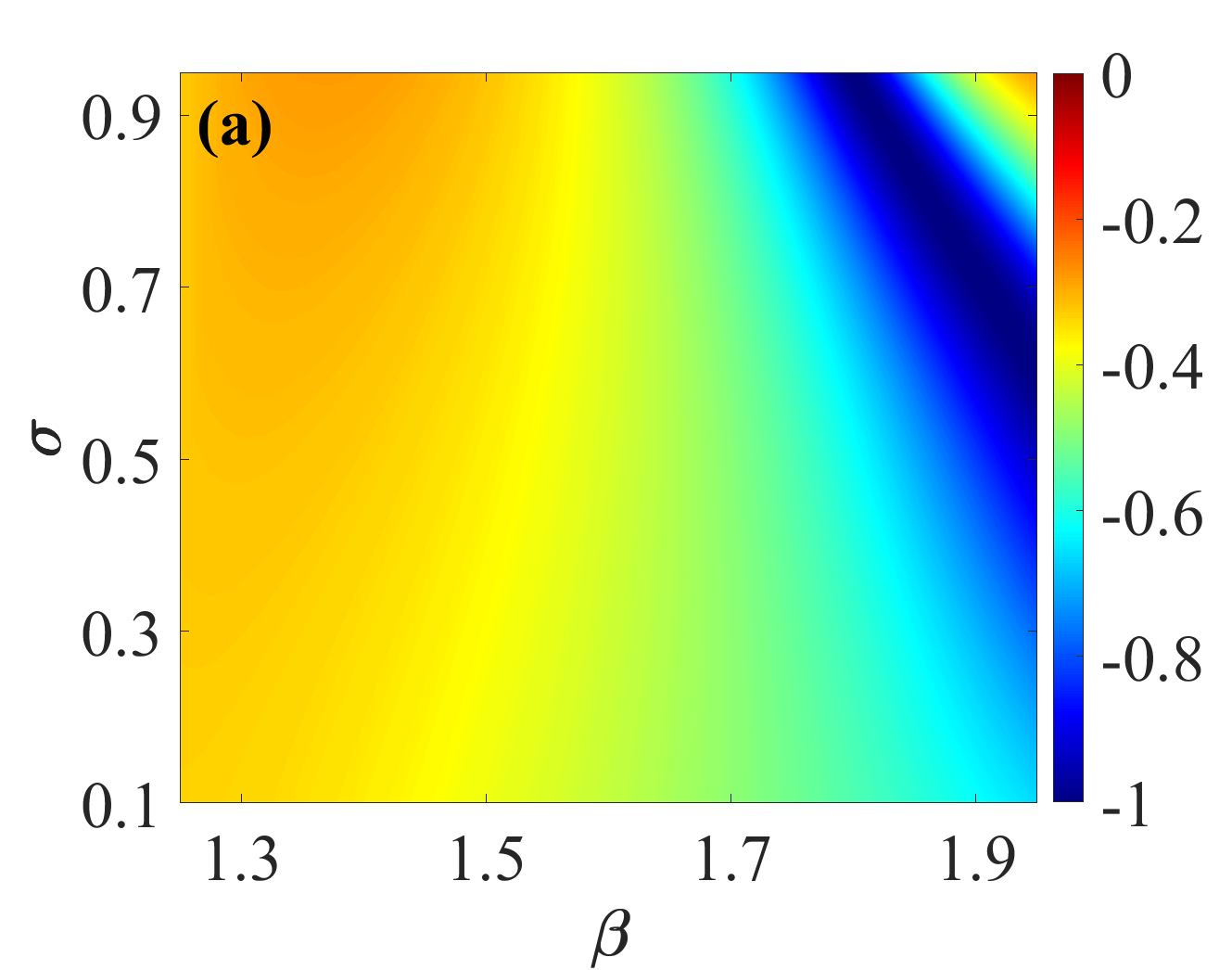}\hfil
\includegraphics[width=0.49\linewidth]{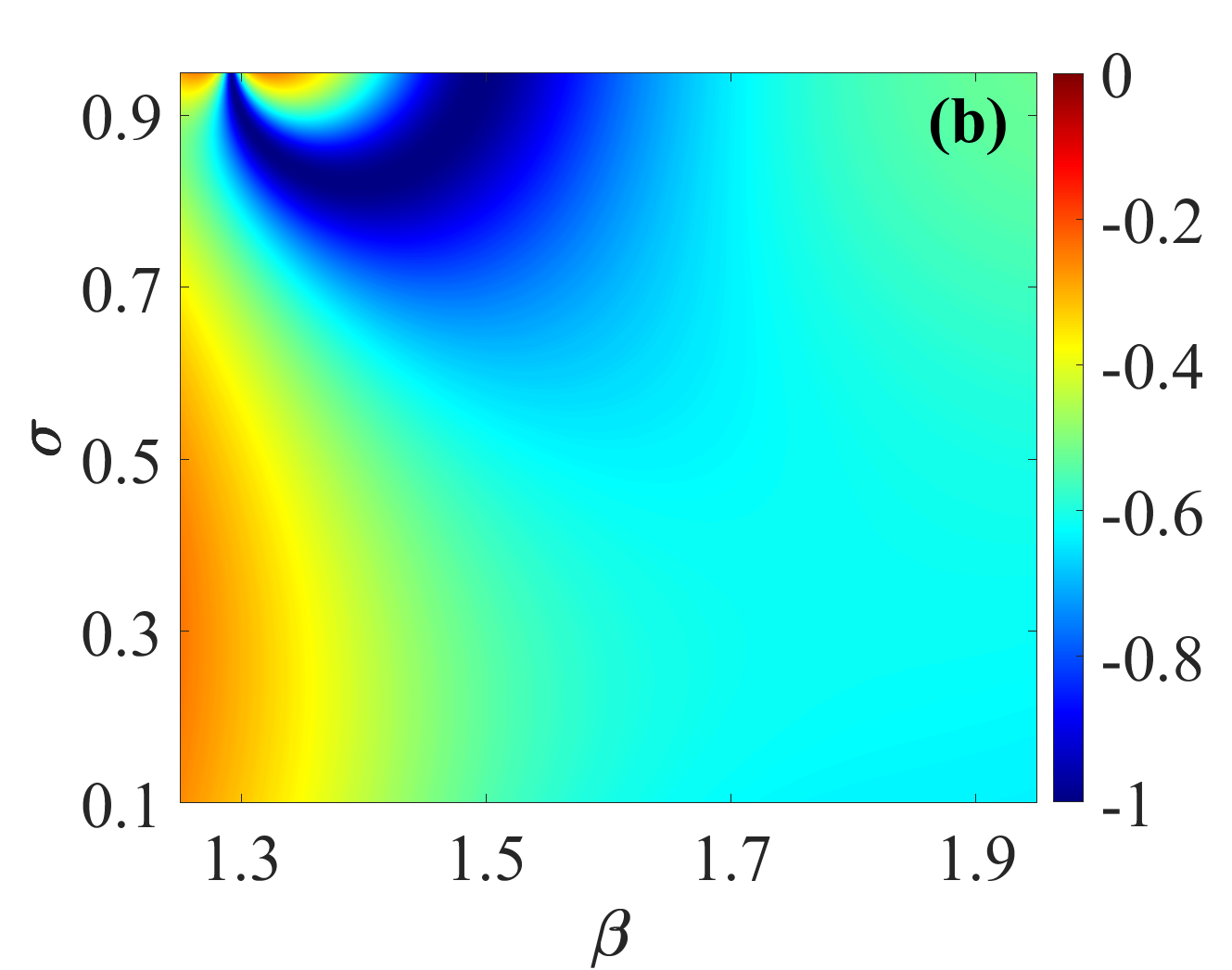}
\caption{Contour map of the supercurrent rectification amplitude $\eta$ for the Josephson junction as a function of the  fractional orders $1<\beta < 2$ and $0<\sigma < 1$ for $\epsilon/\sqrt{|a|\gamma}=0.9$ and for the length $L=5\sqrt{\gamma/\|a|}$ with the pure right-sided kinetic mixing weight $\kappa=0$ and pure left-sided Lifshitz mixing weight $\mu=1$ (a) and with their equal contributions $\kappa=\mu=0.5$ (b).}
\label{diode_eff_plot}
\end{figure*}

Based on Eq. \eqref{eq:CPR-final} one can derive the rectification amplitude $\eta$ in the analytical form:
\begin{equation}
\eta = - \frac{\varepsilon\,|\Delta_0|^2}{\sqrt{J_{\beta}^2+\big(\varepsilon|\Delta_0|^2\big)^2}}\,.
\label{eq:eta-diode}
\end{equation}

Using Eq. \eqref{eq:eta-diode}, one can study the effect of the orders of Caputo fractional derivatives on the amplitude of the rectification coefficient of a given Josephson junction. The results of this numerical study are presented in the form of a phase diagram in Figure \ref{diode_eff_plot} for asymmetric mixing left- and right-sided weights of fractional derivatives $\kappa$ and $\mu$ (Fig. \ref{diode_eff_plot}a) and for their equal contributions (Fig. \ref{diode_eff_plot}b). The phase diagrams clearly indicate the existence of regions (dark blue domains) with pronounced, practically ideal Josephson diode efficiency, corresponding to non-integer orders of Caputo fractional derivatives. We deliberately confined the interval of values $\beta$ and $\sigma$ on the phase diagrams to stress that the limiting case of integer orders of Caputo fractional derivatives was excluded from the numerical calculations.

%fractional derivatives  provide a natural, gauge-consistent way to model nonlocality and inversion asymmetry and to tune the amplitude and scaling of the effect. 

%Thus, in our fractional derivative-based GL approach, nonreciprocity arises from the interplay of the Lifshitz term with the fractional kinetics; setting  $\varepsilon=0$ eliminate the diode effect regardless of the order of $\beta$ and $\sigma$.

Thus, in our fractional GL approach the Josephson diode effect arises from the interplay of two key ingredients: the Lifshitz invariant, which generates the $\varphi_0$ component of the current–phase relation, and the fractional derivative kinetics, which encode nonlocality and inversion asymmetry. While the Lifshitz term is the minimal and indispensable source of diode nonreciprocity, the fractional operators amplify and reshape the effect by modifying the current prefactor and its scaling with Josephson system size. Setting  $\varepsilon=0$ one can eliminate the diode effect regardless of the order of $\beta$ and $\sigma$.

\subsubsection{The single–sided solution with $\kappa=\mu=1$ and the integer order of the Lifshitz invariant $\sigma=1$}

Although in the previous subsection we considered the entire interval of fractional orders of derivatives using an exponential ansatz, it is instructive to supplement our treatment by the limiting case of a linearized one-sided fractional GL equation when $\kappa=\mu=1$, in which the kinetic term is still described by a Caputo fractional derivative $1<\beta \leq 2$, while the Lifshitz invariant is represented by an integer first derivative as in the conventional case. Within this simplification Eq. \eqref{eq:fractional_gl} reduces to the single–sided Caputo differential equation:
\begin{equation}
-\gamma\,{}_{0}^{C}D_x^{\beta}\,\Delta(x)\;+\;i\,\varepsilon\,\Delta'(x)\;+\;a\,\Delta(x)\;=\;0,
\label{eq:caputo-ode}
\end{equation}
and admits the analytical closed–form solution that satisfies the boundary conditions given by Eq. \eqref{eq:BC}
\begin{equation}
\Delta(x)=|\Delta_0|\Biggl[f_0(x)+\frac{e^{i\varphi}-f_0(L)}{f_1(L)}\,f_1(x)\Biggr].
\label{eq:delta-final}
\end{equation}

Here $f_0,f_1$ are functions represented by infinite series of Prabhakar functions (three-parameter Mittag-Leffler functions): 
%\begin{equation}
%\label{eq:f0f1-prabhakar}
%\begin{aligned}
%f_0(x)
%&=\sum_{m=0}^{\infty}\Bigl(\frac{i\varepsilon}{\gamma}\Bigr)^m
%x^{\,m(\beta-1)}\!
%\Biggl[
%E_{\beta,\;1+m(\beta-1)}^{\,m+1}\!\Bigl(\tfrac{a}{\gamma}\,x^{\beta}\Bigr)
%-\frac{i\varepsilon}{\gamma}\,x^{\,\beta-1}\,
%E_{\beta,\;\beta+m(\beta-1)}^{\,m+1}\!\Bigl(\tfrac{a}{\gamma}\,x^{\beta}\Bigr)
%\Biggr],\\[2pt]
%f_1(x)
%&=\sum_{m=0}^{\infty}\Bigl(\frac{i\varepsilon}{\gamma}\Bigr)^m
%x^{\,1+m(\beta-1)}\,
%E_{\beta,\;2+m(\beta-1)}^{\,m+1}\!\Bigl(\tfrac{a}{\gamma}\,x^{\beta}\Bigr).
%\end{aligned}
%\end{equation}
\begin{equation}
\label{eq:f0f1-prabhakar}
\begin{aligned}
f_0(x)
&= \sum_{m=0}^{\infty}
\Bigl(\tfrac{\mathrm{i}\varepsilon}{\gamma}\Bigr)^m
x^{\,m(\beta-1)}
\Biggl[
E_{\beta,\;1+m(\beta-1)}^{\,m+1}
\!\Bigl(\tfrac{a}{\gamma}\,x^{\beta}\Bigr)
\\
&\qquad
-\,\tfrac{\mathrm{i}\varepsilon}{\gamma}\,
x^{\,\beta-1}\,
E_{\beta,\;\beta+m(\beta-1)}^{\,m+1}
\!\Bigl(\tfrac{a}{\gamma}\,x^{\beta}\Bigr)
\Biggr],
\\[6pt]
f_1(x)
&= \sum_{m=0}^{\infty}
\Bigl(\tfrac{\mathrm{i}\varepsilon}{\gamma}\Bigr)^m
x^{\,1+m(\beta-1)}\,
E_{\beta,\;2+m(\beta-1)}^{\,m+1}
\!\Bigl(\tfrac{a}{\gamma}\,x^{\beta}\Bigr).
\end{aligned}
\end{equation}

The Prabhakar series in Eq. \eqref{eq:f0f1-prabhakar} smoothly transform to the standard Mittag–Leffler functions when $\varepsilon\to 0$ and to the exponential representation from Ref. \onlinecite{Buzdin} when $\beta= 2$.

Based on the general solution Eq. \eqref{eq:f0f1-prabhakar} and the expressions for the $\beta$ and $\sigma$ components of the current density Eqs. \eqref{j_beta} and \eqref{j_sigma}, after long but straightforward calculations, one can obtain the current–phase relation:
\begin{equation}
j(\varphi)=J_{\beta}\,\sin\varphi+J_{\sigma}\,\cos\varphi+J_0,
\label{eq:CPR-trig}
\end{equation}
with corresponding coefficients
\begin{align}
J_{\beta} &= \left[-2\gamma\,\Re (A-C)+\varepsilon\,\Im A\right] |\Delta_0|^2,
\label{eq:J_beta}
\end{align}
\begin{align}
J_{\sigma} &= \left[\,2\gamma\,\Im (A-C)+\varepsilon\,\Re A\right] |\Delta_0|^2,
\label{eq:J_sigma}
\end{align}
\begin{align}
J_0 &=\left[\,2\gamma\,\Im(B-D)+\varepsilon\,\Re B\right] |\Delta_0|^2.
\label{eq:J0}
\end{align}
where, for compactness, we introduced four  complex constants  evaluated at $x=L$:
\begin{equation}
\label{eq:ABCD}
\begin{split}
A &= f_0'(L)-f_0(L)\frac{f_1'(L)}{f_1(L)}, \\
B &= \frac{f_1'(L)}{f_1(L)}, \\
C &= -\frac{f_0(L)}{f_1(L)}, \\
D &= \frac{1}{f_1(L)}.
\end{split}
\end{equation}

The structure of the current-phase relation Eq. \eqref{eq:CPR-trig} consists of the odd amplitude in $\varphi$ Josephson phase $J_{\beta}$, the non-Hermitian drift (Lifshitz) part
$J_{\sigma}$, which is the even-in-$\varphi$ harmonic that produces a $\varphi_0$-shift, and $J_0$ is a $\varphi$-independent offset. 

In the integer limit $\beta = 2$, one can show that $J_0 = 0$ and Eq. \eqref{eq:CPR-trig} collapse to a pure shifted sinusoid $j(\varphi)=\mathcal{J_0}\sin(\varphi+\varphi_0)$ , i.e., no diode effect. For $1<\beta<2$, fractality (the fractional order) produces $J_0\neq 0$, yielding a finite diode efficiency.

In the absence of non-Hermitian Lifshitz drift $\varepsilon=0$ Eqs. \eqref{eq:CPR-trig}–\eqref{eq:J0} reduce to the purely sinusoidal current-phase relation with $J_{\sigma}=J_0=0$ and $j(\varphi)=J_{\beta}\sin\varphi$.

\section{Conclusions}
We have presented a fractional-order GL description of nonreciprocal superconducting transport in Josephson junctions that couples fractal (or otherwise nonlocally correlated) superconductors to a noncentrosymmetric normal layer. Using two complementary approaches - a fractional integral  (measure-weighted) GL theory and a derivative-based GL equations with left/right Caputo operators derived via Agrawal variational principle we have obtained nonreciprocal current–phase relations for a Josephson junction. A two-mode plane-wave approximation and the exact single-sided solution in terms of generalized Mittag–Leffler (Prabhakar) functions of the fractional GL equation show that the superconducting diode effect emerges from the interplay between the Lifshitz invariant and fractional kinetics acting both as an amplifier and shaper of the superconducting nonreciprocal transport.
A central outcome is tunability: by varying the effective fractal dimensionality (captured either by the fractal measure or by the fractional orders controlling kinetic and non-Hermitian drift terms) and the strength/sign of the Lifshitz invariant, the diode efficiency spans nearly the full range, with both strong rectification and near-ideal diode response. In the integer limit, the current-phase relation reduces to the standard phase-shifted $\varphi_0$-junction, recovering the conventional result within the same GL framework.
Our results identify a possible route to near-ideal superconducting diodes in Josephson junctions by engineering fractal (fractional-kinetic) transport, without relying on magnetic fields or geometric asymmetry.

\section*{Acknowledgments}
Y. Y. acknowledges the funding received from the program TÜBİTAK-2221 Fellowships for Visiting Scientists and Scientists on Sabbatical Leave and from HPC National Center for HPC, Big Data and Quantum Computing - HPC (Centro Nazionale 01 – CN0000013). 

%\newpage

\appendix
\begin{widetext}

\section{Fourier analysis of current-phase relations in a fractional integral GL formulation}
\label{App_A}

The cumbersome analytical expression for the current-phase relation obtained within the framework of the unweighted GL formalism does not allow us to analyze its phase structure. Therefore, the best method for understanding the harmonic components of the dependence $j(\varphi)$ is Fourier analysis.

The periodic function representing the current-phase relation for a given Josephson junction $I_j (\phi)$ can be generally expanded into Fourier series:
\begin{equation}
 j(\varphi) = a_0 + \sum_{n=1}^{\infty} \left[ a_n \sin n\varphi + b_n \cos n\varphi \right],
 \label{eq:full_fourier}
\end{equation}
where the coefficients $a_n$, $b_n$, and $a_0$ are given by:
\begin{align}
a_0 &= \frac{1}{2\pi} \int_{-\pi}^{\pi} j(\varphi)\, d\varphi, \\
a_n &= \frac{1}{\pi} \int_{-\pi}^{\pi} j(\varphi) \sin n\varphi \, d\varphi, \\
b_n &= \frac{1}{\pi} \int_{-\pi}^{\pi} j(\varphi) \cos n\varphi \, d\varphi.
\end{align}

Each term in the decomposition of Eq.~\eqref{eq:full_fourier} has a distinct physical meaning, namely $a_1$ is the leading odd-parity harmonic. In Josephson junctions, $a_n$ with $n > 1$ are higher-order odd-parirty harmonics, which may arise from multiband effects, strong coupling, or higher-order tunneling processes. As expected, for sizable values of $b_2$, $b_3$, etc., the behavior of the current phase relation significantly deviates from the standard sinusoidal Josephson profile. In turn, the coefficients $b_n$ correspond to even parity cosine harmonics, which are typically zero in the presence of time-reversal symmetry or when the parity of the current-phase relation is an odd function of $\varphi$. Their presence implies time-reversal symmetry breaking thus yielding $\pi$ or $\varphi_0$ Josephson junctions. The first term $a_0$ is the dc component,  which manifests a net supercurrent even at zero phase difference, signaling broken inversion and/or time-reversal symmetry.

\begin{figure*}[h]
\centering
\includegraphics[width=0.49\columnwidth]{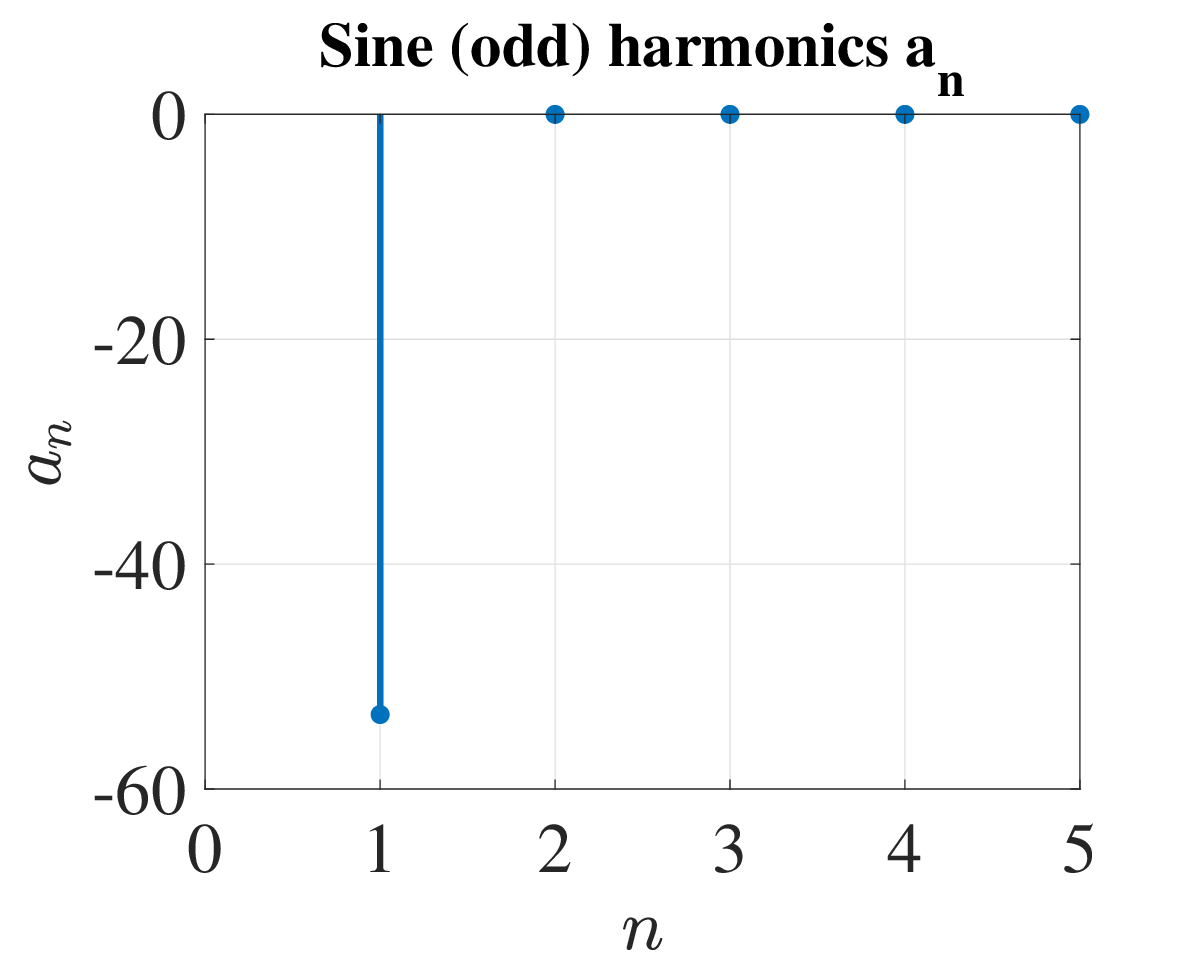}
\includegraphics[width=0.49\columnwidth]{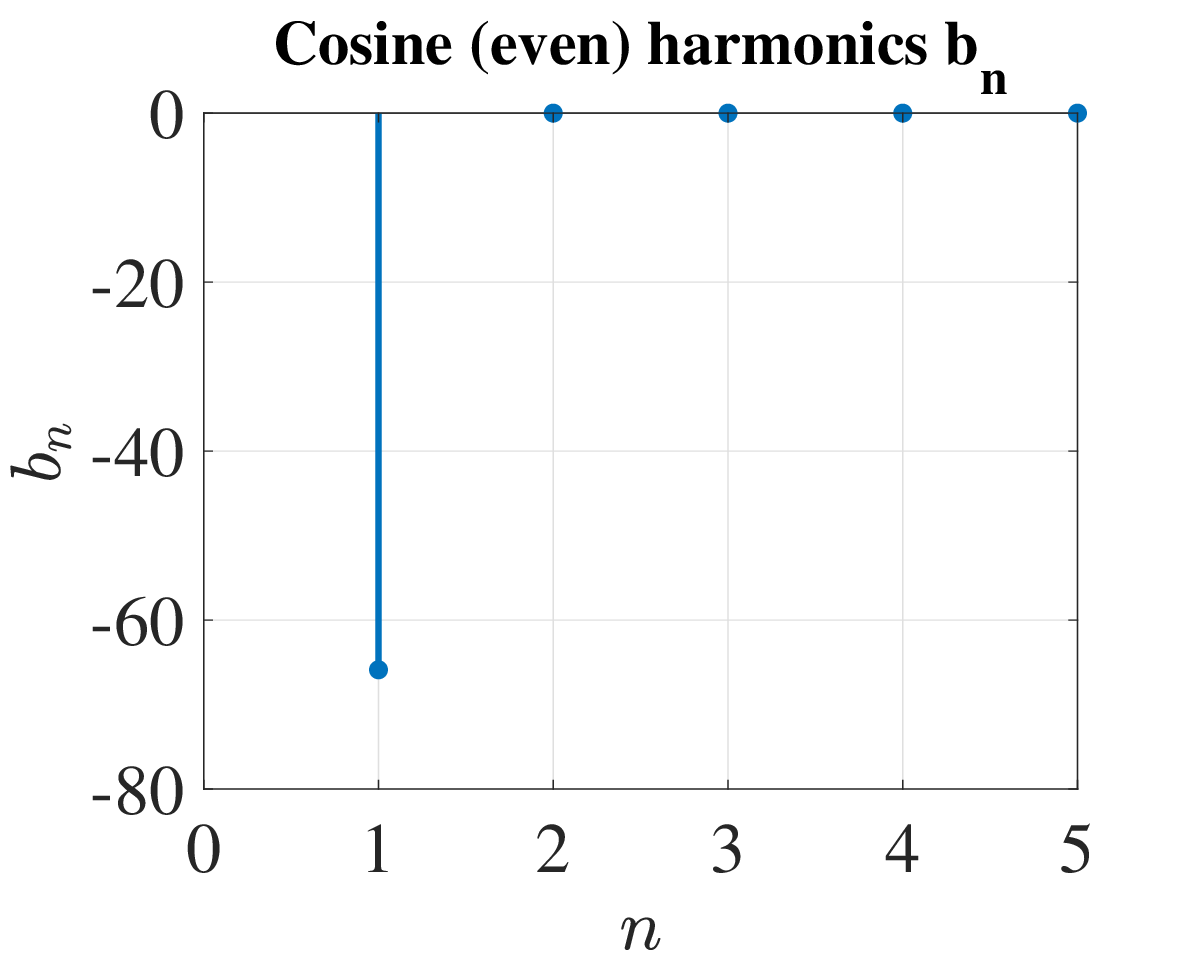}
\caption {Fourier analysis of the current-phase relation shown in Figure \ref{CPR_frac_int} displaying the first five harmonics $a_n$, $b_n$ for the case when the diode efficiency $\eta \approx -0.869$.}
\label{Fourier_current}
\end{figure*}

The Fourier analysis visualized in Figure \ref{Fourier_current}  clearly indicates the presence of the first harmonics only $n=1$ with the nonzero offset value $a_0$. Consequently, the current-phase relation of the Josephson junction calculated within the "unweighted" formulation of the GL theory can indeed be represented effectively by Eq. \eqref{CPR_unweighted} as shown in the main text, where $a_1 \equiv \mathcal{A}$ $b_1 \equiv \mathcal{B}$ and $a_0 \equiv \mathcal{C}$.

\section{Fractional GL functional and current density from Agrawal’s variational principle}
\label{App_B}

\subsection{Setting, operators, and fractional integration by parts}

We recall that the left/right Caputo derivatives of order
\(0<\nu\le 1\) and \(1<\nu\le 2\) are (with \(n-1<\nu\le n\), \(n\in\{1,2\}\)) are defined as
\begin{equation}
{}_{0}^{C}\!D_{x}^{\nu} f(x)
=\frac{1}{\Gamma(n-\nu)}\!\int_{0}^{x}\!\frac{f^{(n)}(\xi)}{(x-\xi)^{\nu-n+1}}\,d\xi,\qquad
{}_{x}^{C}\!D_{L}^{\nu} f(x)
=\frac{(-1)^{n}}{\Gamma(n-\nu)}\!\int_{x}^{L}\!\frac{f^{(n)}(\xi)}{(\xi-x)^{\nu-n+1}}\,d\xi.
\end{equation}
The Riemann–Liouville fractional integrals are introduced as
\begin{equation}
{}_{0}I_{x}^{\mu} f(x)=\frac{1}{\Gamma(\mu)}\!\int_{0}^{x}\!(x-\xi)^{\mu-1}f(\xi)\,d\xi,\qquad
{}_{x}I_{L}^{\mu} f(x)=\frac{1}{\Gamma(\mu)}\!\int_{x}^{L}\!(\xi-x)^{\mu-1}f(\xi)\,d\xi,\quad \mu>0.
\end{equation}

We use the following two identities (valid for sufficiently smooth functions; see, e.g., \cite{SamkoKilbasMarichev,Agrawal}):
\begin{align}
\int_{0}^{L}\!\phi(x)\,\bigl({}_{0}^{C}\!D_{x}^{\sigma}\psi(x)\bigr)\,dx
&= \Bigl[\phi\,\, {}_{0}I_{x}^{\,1-\sigma}\psi\Bigr]_{0}^{L}
+\int_{0}^{L}\!\bigl({}_{x}D_{L}^{\sigma}\phi\bigr)\,\psi\,dx,
\qquad 0<\sigma\le 1,
\label{eq:FIBP-sigma}\\[4pt]
\int_{0}^{L}\!\phi(x)\,\bigl({}_{0}^{C}\!D_{x}^{\beta}\psi(x)\bigr)\,dx
&= \Bigl[\phi\,\, {}_{0}I_{x}^{\,2-\beta}\psi'
-\phi'\,\, {}_{0}I_{x}^{\,2-\beta}\psi\Bigr]_{0}^{L}
+\int_{0}^{L}\!\bigl({}_{x}D_{L}^{\beta}\phi\bigr)\,\psi\,dx,
\qquad 1<\beta\le 2.
\label{eq:FIBP-beta}
\end{align}
Here ${}_{x}D_{L}^{\nu}$ denotes the right Riemann–Liouville derivative. For later calculations, we also apply the Caputo-Riemann-Liouville identities  on the interval $x \in \left[ {0,L} \right]$, corresponding to the Josephson junction geometry (see Fig. \ref{model}):
\begin{align}
{}_{0}I_{x}^{\,1-\sigma}\,{}_{0}^{C}\!D_{x}^{\sigma} f &= f(x)-f(0),&
{}_{x}I_{L}^{\,1-\sigma}\,{}_{x}^{C}\!D_{L}^{\sigma} f &= -\,f(x)+f(L),& 0<\sigma\le 1,
\label{eq:undo-sigma}\\
{}_{0}I_{x}^{\,2-\beta}\,{}_{0}^{C}\!D_{x}^{\beta} f &= f'(x)-f'(0),&
{}_{x}I_{L}^{\,2-\beta}\,{}_{x}^{C}\!D_{L}^{\beta} f &= -\,f'(x)+f'(L),& 1<\beta\le 2.
\label{eq:undo-beta}
\end{align}

Throughout Appendix \ref{App_B} we treat $\Delta$ and $\Delta^*$ as independent functions. For a Josephson junction we impose Dirichlet boundary conditions from Eq. \eqref{eq:BC}, so that $\delta\Delta(0)=\delta\Delta(L)=0$ and no condition is imposed on their derivatives $\Delta'(0),\Delta'(L)$.

\subsection{Fractional GL functional}

We seek a real quadratic functional \(\mathcal{F}[\Delta]\) whose Euler–Lagrange equation (variation w.r.t. \(\Delta^*\)) reproduces the linear fractional
GL equation (with \(0<\sigma\le 1,\;1<\beta < 2\), \(0\le \kappa,\mu\le 1\)) Eq. \eqref{eq:fractional_gl}.

The corresponding Lagrangian can be written as follows:
\begin{align}
\mathcal{L}[\Delta,\Delta^*]
&= \frac{\gamma}{2}\Big\{
\kappa\Big[\Delta^*\,({}_{0}I_{x}^{\,2-\beta}\,{}_{0}^{C}\!D_{x}^{\beta}\Delta)
+\Delta\,({}_{0}I_{x}^{\,2-\beta}\,{}_{0}^{C}\!D_{x}^{\beta}\Delta)^*\Big]
-(1-\kappa)\Big[\Delta^*\,({}_{x}I_{L}^{\,2-\beta}\,{}_{x}^{C}\!D_{L}^{\beta}\Delta)
+\Delta\,({}_{x}I_{L}^{\,2-\beta}\,{}_{x}^{C}\!D_{L}^{\beta}\Delta)^*\Big]\Big\}
\notag\\
&\quad
+\frac{i\varepsilon}{2}\Big\{
\mu\Big[\Delta^*\,({}_{0}I_{x}^{\,1-\sigma}\,{}_{0}^{C}\!D_{x}^{\sigma}\Delta)
-\Delta\,({}_{0}I_{x}^{\,1-\sigma}\,{}_{0}^{C}\!D_{x}^{\sigma}\Delta)^*\Big]
-(1-\mu)\Big[\Delta^*\,({}_{x}I_{L}^{\,1-\sigma}\,{}_{x}^{C}\!D_{L}^{\sigma}\Delta)
-\Delta\,({}_{x}I_{L}^{\,1-\sigma}\,{}_{x}^{C}\!D_{L}^{\sigma}\Delta)^*\Big]\Big\}
\notag\\
&\quad + a\,|\Delta|^{2},
\label{eq:Lag-density}
\end{align}
where the GL free energy is determined as $\mathcal{F}[\Delta]=\int_{0}^{L}\mathcal{L}\,dx$.

\subsection{Euler–Lagrange equation from Agrawal variational principle}

Using Eq. \eqref{eq:FIBP-sigma} with \(\phi=\delta\Delta^*\) and
\(\psi={}_{0}I_{x}^{\,1-\sigma}\,{}_{0}^{C}\!D_{x}^{\sigma}\Delta\),
and the analogous right-sided relation, one finds (boundary terms vanish
because \(\delta\Delta(0)=\delta\Delta(L)=0\))
\begin{equation}
\delta\!\int_{0}^{L}\!\frac{i\varepsilon}{2}\Big\{\eta\big[\Delta^*\,({}_{0}I_{x}^{\,1-\sigma}\,{}_{0}^{C}\!D_{x}^{\sigma}\Delta)\big]
-(1-\eta)\big[\Delta^*\,({}_{x}I_{L}^{\,1-\sigma}\,{}_{x}^{C}\!D_{L}^{\sigma}\Delta)\big]\Big\}dx
= \int_{0}^{L}\! \frac{i\varepsilon}{2}\Big[\eta\,{}_{x}D_{L}^{\sigma}
-(1-\eta)\,{}_{0}D_{x}^{\sigma}\Big]\!\Big({\cdots}\Big)\,\delta\Delta^*\,dx,
\end{equation}
and then, by the Caputo–Riemann-Liouville adjoint property
${}_{x}D_{L}^{\sigma}({}_{0}I_{x}^{\,1-\sigma}\,{}_{0}^{C}\!D_{x}^{\sigma}\Delta)
={}_{0}^{C}\!D_{x}^{\sigma}\Delta$ and
${}_{0}D_{x}^{\sigma}({}_{x}I_{L}^{\,1-\sigma}\,{}_{x}^{C}\!D_{L}^{\sigma}\Delta)
={}_{x}^{C}\!D_{L}^{\sigma}\Delta$
(up to endpoint terms that vanish under the boundary conditions), yielding the desired
combination $i\varepsilon\big[\eta\,{}_{0}^{C}\!D_{x}^{\sigma}
+(1-\eta)\,{}_{x}^{C}\!D_{L}^{\sigma}\big]\Delta$.

Using Eq. \eqref{eq:FIBP-beta} with
\(\phi=\delta\Delta^*\) and
\(\psi={}_{0}I_{x}^{\,2-\beta}\,{}_{0}^{C}\!D_{x}^{\beta}\Delta\)
(and similarly for the right-sided term), boundary contributions again drop out
because they are proportional to \(\delta\Delta\) evaluated at the endpoints.
The volume term becomes
\[
\int_{0}^{L}\!\frac{\gamma}{2}\Big[\kappa\,{}_{x}D_{L}^{\beta}
-(1-\kappa)\,{}_{0}D_{x}^{\beta}\Big]\!\Big({\cdots}\Big)\,\delta\Delta^*\,dx
\;\xrightarrow\;
-\gamma\Big[\kappa\,{}_{0}^{C}\!D_{x}^{\beta}+(1-\kappa)\,{}_{x}^{C}\!D_{L}^{\beta}\Big]\Delta,
\]
where we use that
\({}_{x}D_{L}^{\beta}({}_{0}I_{x}^{\,2-\beta}\,{}_{0}^{C}\!D_{x}^{\beta}\Delta)
={}_{0}^{C}\!D_{x}^{\beta}\Delta\) and
\({}_{0}D_{x}^{\beta}({}_{x}I_{L}^{\,2-\beta}\,{}_{x}^{C}\!D_{L}^{\beta}\Delta)
={}_{x}^{C}\!D_{L}^{\beta}\Delta\)
again up to endpoint terms canceled by our Dirichlet variation.

Finally, the potential term gives \(\delta\!\int\alpha|\Delta|^{2}\,dx
=\int \alpha\,\Delta\,\delta\Delta^{*}\,dx\). Collecting all pieces,
the Euler–Lagrange equation $\delta\mathcal{F}/\delta\Delta^{*}=0$ reproduces
Eq.~\eqref{eq:fractional_gl}:
\begin{equation}
-\gamma\!\left[\kappa\,{}_{0}^{C}\!D_{x}^{\beta}\Delta
+(1-\kappa)\,{}_{x}^{C}\!D_{L}^{\beta}\Delta\right]
+i\varepsilon\!\left[\eta\,{}_{0}^{C}\!D_{x}^{\sigma}\Delta
+(1-\eta)\,{}_{x}^{C}\!D_{L}^{\sigma}\Delta\right]
+a\,\Delta=0.
\end{equation}

\subsection{The current density expression}

The GL functional $\mathcal{F}$ is invariant under a global phase rotation
$\Delta\to e^{i\theta}\Delta$ ($\theta$ constant). The associated current density follows from the first variation with $\delta\Delta=i\theta\,\Delta$, $\delta\Delta^{*}=-i\theta\,\Delta^{*}$, with only \(O(\theta)\). A standard variational calculus procedure yields a current density expression $j(x)$
\begin{equation}
j(x)=j_{\beta}(x)+j_{\sigma}(x),
\qquad
j_{\beta}(x)=2\gamma\,\Im\!\Big\{\Delta^{*}(x)\,\mathcal{K}_{\beta}\,\Delta(x)\Big\},
\qquad
j_{\sigma}(x)=\varepsilon\,\Re\!\Big\{\Delta^{*}(x)\,\mathcal{D}_{\sigma}\,\Delta(x)\Big\},
\label{eq:j-total}
\end{equation}
where the Caputo-Riemann-Liouville identities are
\begin{align}
\mathcal{K}_{\beta}
&:=\kappa\,{}_{0}I_{x}^{\,2-\beta}\,{}_{0}^{C}\!D_{x}^{\beta}
-(1-\kappa)\,{}_{x}I_{L}^{\,2-\beta}\,{}_{x}^{C}\!D_{L}^{\beta},
\qquad 1<\beta\le 2, \label{eq:Kbeta}\\[2pt]
\mathcal{D}_{\sigma}
&:=\mu\,{}_{0}I_{x}^{\,1-\sigma}\,{}_{0}^{C}\!D_{x}^{\sigma}
-(1-\mu)\,{}_{x}I_{L}^{\,1-\sigma}\,{}_{x}^{C}\!D_{L}^{\sigma},
\qquad 0<\sigma\le 1. \label{eq:Dsigma}
\end{align}
Eqs. \eqref{eq:undo-sigma}–\eqref{eq:undo-beta} show that
\(\mathcal{D}_{\sigma}\) and \(\mathcal{K}_{\beta}\) localize nonlocal Caputo derivatives in endpoint–subtracted expressions. For example,
\begin{equation}
\mathcal{D}_{\sigma}\Delta
=\mu\,[\Delta(x)-\Delta(0)]-(1-\mu)\,[\Delta(x)-\Delta(L)]
\end{equation},
and
\begin{equation}
\mathcal{K}_{\beta}\Delta
=\eta_{\beta}\bigl[\Delta'(x)-\Delta'(0)\bigr]
-(1-\eta_{\beta})\bigl[\Delta'(x)-\Delta'(L)\bigr]
\end{equation}
In particular, the non-Hermitian drift/Lifshitz term $j_{\sigma}$ produces the $\cos\varphi$ contribution to the current-phase relation under Josephson boundary conditions Eqs. \eqref{eq:BC}, while $j_{\beta}$ gives the odd $\sin\varphi$ harmonic via the kinetic overlap (see the main text).

In the local GL limit for $\beta= 2$, $\sigma = 1$, the corresponding relations reduce to
${}_{0}I_{x}^{\,2-\beta}\to{}_{0}I_{x}^{0}=\mathrm{Id}$, 
${}_{x}I_{L}^{\,2-\beta}\to \mathrm{Id}$,
${}_{0}I_{x}^{\,1-\sigma}\to \mathrm{Id}$,
${}_{x}I_{L}^{\,1-\sigma}\to \mathrm{Id}$, i.e., the fractional integrals of order $0$ just return the same function unchanged. Therefore, the current
\eqref{eq:j-total} collapses to the standard local expressions
\begin{equation}
j_{\beta}\to 2\gamma\,\Im\{\Delta^{*}\,\partial_{x}\Delta\},\qquad
j_{\sigma}\to =-\varepsilon\,(2\mu-1)\,|\Delta|^{2},
\end{equation}
which, upon combining the odd/even harmonics, give rise a single $\phi_{0}$–shifted sinusoid for the Josephson current-phase relation without diode effect .

\section{Consistency check of the fractional GL theory for a $\varphi_0$-Josephson junction}
\label{App_C}

If we put $\beta = 2$ and $\sigma = 1$ in the mixed left–right Caputo GL model on $[0,L]$, and keep a fixed mixing $\kappa\in[0,1]$ in the kinetic channel and $\mu\in[0,1]$ in the non-Hermitian drift channel, then the fractional GL equation and the current density reduce to the forms
\begin{equation}
-\gamma\,\Delta''(x)+i\,\varepsilon_{\rm eff}\,\Delta'(x)+\alpha\,\Delta(x)=0,
\qquad
j(x)=2\gamma\,\Im\{\Delta^*(x)\Delta'(x)\}-\varepsilon_{\rm eff}\,|\Delta(x)|^2,
\end{equation}
with the effective Lifshitz invariant coefficient
\begin{equation}
\varepsilon_{\rm eff} \;=\; (2\mu-1)\,\varepsilon.
\end{equation}

Hence, for a Josephson weak link of length $L$ , the solution produces a well-known shifted current–phase relation $j(\varphi)=j_c\sin(\varphi+\varphi_0)$ with phase shift:
\begin{equation}
\varphi_0=\frac{\varepsilon_{\rm eff}}{2\gamma}\,L.
\end{equation}
In particular, for geometry of Ref. \onlinecite{Buzdin} $L$ is substituted by $2L$ and the identification $\varepsilon_{\rm eff}=2\,\varepsilon_{\rm so}h$ one recovers $\varphi_0=2(\varepsilon_{\rm so}h/\gamma)L$.

\end{widetext}

\bibliography{REF}

\end{document}